\documentclass[fleqn,usenatbib]{mnras}

\usepackage{mathptmx}

\usepackage[T1]{fontenc}
\usepackage{ae,aecompl}

\usepackage{graphicx}	
\usepackage{amsmath}	
\usepackage{amssymb}	
\usepackage{comment}
\usepackage{caption}
\usepackage{subcaption}

\usepackage{pdflscape}
\usepackage{hyperref}
\usepackage{microtype}
\usepackage{color}
\usepackage[utf8]{inputenc}

\newcommand{\HI}{H$\,${\sc i}}

\newcommand{\nhi}{$N_{\rm H\,I}$}

\newcommand{\lya}{Ly$\alpha$}

\defcitealias{Khaire23}{Paper I}

\graphicspath{{./}{figures/}}

\title[Feedback and Ly$\alpha$ Forest II]{\huge
Searching for the Imprints of AGN Feedback on the Lyman Alpha
Forest Around Luminous Red Galaxies}
\author[Khaire et al.]
{
\parbox{\textwidth}{
Vikram Khaire,$^{1,2}$
Teng Hu,$^2$ Joseph F. Hennawi,$^{2, 3}$ Joseph N. Burchett,$^{4}$ \\
Michael Walther$^{5, 6}$ and Frederick Davies$^{\,2,7}$ 
} 
\vspace*{10pt}\\
$^{1}$Indian Institute of Space Science \& Technology, Thiruvananthapuram, Kerala  695547, India\\
$^{2}$Physics Department, Broida Hall, University of California Santa Barbara, Santa Barbara, CA 93106-9530, USA\\
$^{3}$Leiden Observatory, Leiden University, PO Box 9513, NL-2300 RA Leiden, the Netherlands\\
$^{4}$Department of Astronomy, New Mexico State University, 1320 Frenger Mall, Las Cruces, NM 88003-8001, USA\\
$^{5}$University Observatory, Faculty of Physics, Ludwig-Maximilians-Universit\"{a}t M\"{u}nchen, Scheinerstr. 1, 81677 M\"{u}nchen, Germany\\
$^{6}$Excellence Cluster ORIGINS, Boltzmannstrasse 2, D-85748 Garching, Germany\\
$^{7}$Max-Planck-Institut für Astronomie, Königstuhl 17, 69117 Heidelberg, Germany\\
} 
 
\pubyear{2023}
\begin{document}
\label{firstpage}
\pagerange{\pageref{firstpage}--\pageref{lastpage}}
\maketitle





\begin{abstract}
We explore the potential of using the low-redshift Lyman-$\alpha$ (Ly$\alpha$) forest surrounding
luminous red galaxies (LRGs) as a tool to constrain active galactic nuclei (AGN) 
feedback models. Our analysis is based on snapshots from the Illustris and IllustrisTNG 
simulations at a redshift of 
$z=0.1$. These simulations offer an ideal platform for studying the influence of AGN feedback 
on the gas surrounding galaxies, as they share the same initial conditions and underlying code 
but incorporate 
different feedback prescriptions.
Both simulations show significant impacts of feedback on the temperature and density of 
the gas around massive halos. Following our previous work, we adjusted the UV background in 
both simulations to align with the observed number density of Ly$\alpha$ lines ($\rm dN/dz$) in 
the intergalactic medium and  study the Ly$\alpha$ forest around massive halos hosting 
LRGs, at impact parameters ($r_{\perp}$) ranging from 0.1 to 100 pMpc.
Our findings reveal that $\rm dN/dz$, as a function of $r_{\perp}$, is 
approximately 1.5 to 2 times higher in IllustrisTNG compared to Illustris up to 
$r_{\perp}$ of $\sim 10$ pMpc. To further assess whether existing data can effectively 
discern these differences, we search for archival data containing spectra of background quasars 
probing foreground LRGs. Through a feasibility analysis based on this data, 
we demonstrate that  ${\rm dN/dz} (r_{\perp})$ measurements can distinguish between 
feedback models of IllustrisTNG and Illustris with a precision exceeding 12$\sigma$. 
This underscores the potential of ${\rm dN/dz} (r_{\perp})$ measurements around LRGs 
as a valuable benchmark observation for discriminating between different feedback models.
\end{abstract}

\begin{keywords}
{Add these before submission}
\end{keywords}
\section{Introduction} \label{sec:intro}
The formation and evolution of galaxies remain some of the most intriguing enigmas in 
modern astrophysics. Among the key puzzles is the disparity between blue star-forming 
galaxies and red quiescent elliptical galaxies, the steep decline of cosmic star 
formation rate density at $z<2$, as well as the suppression of luminous 
galaxies in massive dark matter halos. The underlying physics behind these phenomena and 
the mechanisms responsible for shaping the fate of galaxies are still open 
questions in the field of galaxy formation \citep[see reviews by][]{Naab17, Vogelsberger20}. 
Current cosmological simulations attempt to reproduce these observations by harnessing 
the power of central supermassive black holes, by a mechanism generally referred to as 
Active Galactic Nuclei (AGN) feedback, which drives galactic-scale winds that stifle star 
formation in massive galaxies at late cosmic times. However, the exact manifestation of 
AGN feedback is hotly debated \citep[see e.g][]{Springel05, Croton06, Sijacki07, Hopkins08},
and the diversity of modeling approaches in recent simulations 
\citep[e.g][]{Horizon_agn, Illustris, Eagle, massive_balckII, Sherwood, IllustrisTNG, Simba} 
are a testament to the limited understanding and observational 
constraints of this intricate physical process.

Cosmological simulations of galaxy formation adjust their feedback parameters 
using various observations such as  star
formation rate density, stellar mass function, galaxy sizes and
color distribution, and total gas fraction within a radius 
$R_{500c}$\footnote{$R_{500c}$ represents the radius of a sphere surrounding halos, 
encompassing a mass density that is 500 times the critical matter density.}
of halo \citep[as done in IllustrisTNG;][]{IllustrisTNG}. 
But there are no simulations that adjust the parameters using observations
of gas around galaxies beyond $R_{500c}$ and into the intergalactic medium (IGM). 
Because these AGN feedback driven powerful 
galactic-scale winds and outflows
not only stifle star formation in galaxies but also influence the gas in the circumgalactic medium (CGM) and the IGM \citep[][]{Gurvich17, Burkhart22, Tilman23, Mallik23, Khaire23}, 
it would be important to investigate if the observations
of CGM and IGM can be used to constrain the parameters of feedback.

With this objective in mind, in our recent work 
\citep[][hereafter \citetalias{Khaire23}]{Khaire23}, we studied the IGM in Illustris 
\citep{Illustris} and IllustrisTNG \citep{IllustrisTNG} simulations at $z<0.3$. 
These simulations provide an ideal test-bed for 
studying the effect of AGN feedback on CGM and IGM. This is because both simulations 
share the same underlying code and similar cosmological parameters and 
the primary difference between them lies in their 
implementation of AGN feedback \citep[see][]{Pillepich18}. 
Illustris uses a strong radio-mode thermal feedback 
and IllustrisTNG employs a milder kinetic radio-mode feedback. This difference in the 
feedback implementation leads 
to significant differences in the distribution of baryons around massive halos and 
in the IGM. In \citetalias{Khaire23}, we found that feedback can severely affect IGM and 
thereby affect various statistical measures used to study IGM. This is mainly 
because the fraction of cool brayons responsible for the \lya~forset itself is 
affected by the feedback. However, while comparing various IGM statistics with
observations, one needs to adjust the UV background \citep[see][]{HM12, KS19, FG20} in simulations because
the mean optical depth of the \lya~forest depends on the degenerate combination
of the UV background and cool baryon fraction. After such adjustment most of the
statistics of \lya~forest such as line density and 2D
and marginalized distributions of Doppler widths ($b$) and  H~{\sc i} column density ($N_{\rm HI}$), 
do not show any significant difference between simulations. 
Only the \lya~flux power spectrum \citep[as measured in][]{Khaire19} at small 
spatial scales exhibits potentially observable differences 
\citepalias[refer to][for more details]{Khaire23}. 
Therefore it is important to investigate if there is any other
statistic of IGM and CGM that can distinguish the effect of feedback. 

\begin{figure*}
\includegraphics[width=0.99\textwidth,keepaspectratio]{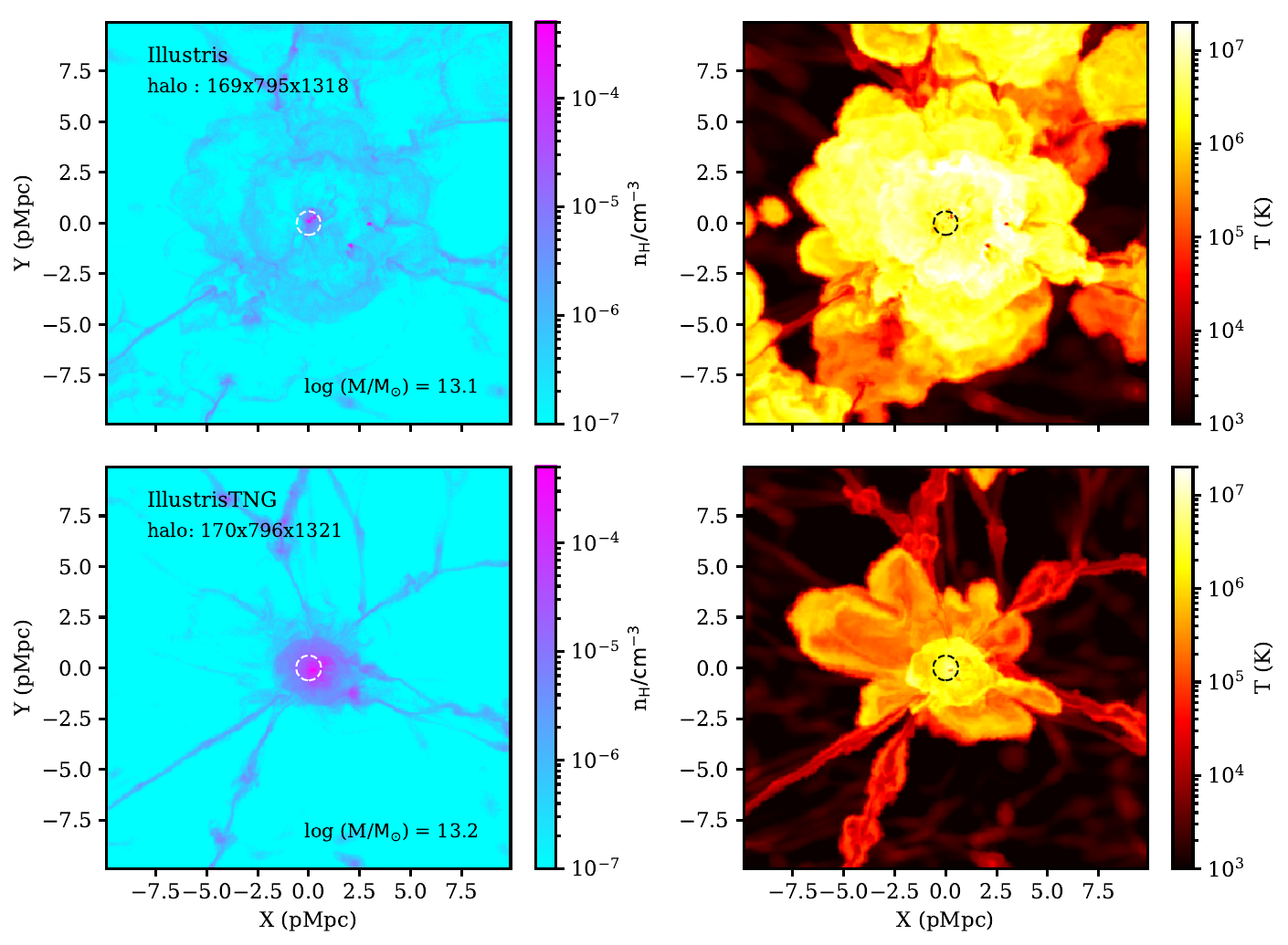}
\caption{
  The figure shows the density (left) and temperature (right) of gas
  around the same halo at $z=0.1$ in Illustris (top ) and IllustrisTNG (bottom)
  simulation. Each panel represents a 2D slice of the simulation (with single cell width 
  i.e $\sim 60$ ckpc) centered at the position of the halo. The explosive
  radio-mode AGN feedback in the Illustris is responsible for displacing the hot gas
  to a large volume as compared to IllustrisTNG. 
  Dashed circles at the center show the Virial sizes of halos. Both 
  halos are identified by matching the coordinates of their location (as noted on the left-hand panels) from halo catalogs. Halo masses are noted in the legends. }
\label{fig.boxplot}
\end{figure*}

Although the IGM's usefulness is somewhat limited due to its degeneracy with the 
UV background, notable distinctions in gas distribution around massive galaxy halos still 
persist (as illustrated in Fig.~\ref{fig.boxplot}). This forms the core of our investigation in 
this paper. Here we study the gas around massive halos as probed by absorption spectra of the 
background quasar. We generate realistic \lya~forest spectra of 
background quasars probing the gas around massive halos in the foreground at different transverse
distances (i.e impact parameter) 
from halos and study a few statistics such as marginalized and  
2D distribution of $b$ and $N_{\rm HI}$, and number density of lines 
as a function of impact parameter, $\rm dN/dz$ ($r_{\perp}$). 
We found that in realistic spectra where we forward model 
the Hubble Space Telescope (HST) Cosmic Origin Spectrograph (COS) data,  
the differences in the $b$-$N_{\rm HI}$ distribution wash out but the number 
of absorption lines vary dramatically in both simulations.
This discrepancy manifests as a significant difference in $\rm dN/dz$ ($r_{\perp}$) 
values of Illustris and IllustrisTNG, extending up to an impact parameter 
of approximately $r_{\perp} \sim 10$ pMpc.

To investigate the utility of  $\rm dN/dz$ ($r_{\perp}$) for distinguishing different feedback 
models as employed in Illustris and IllustrisTNG, we conducted a search for a catalog of quasar-galaxy pairs 
within the available HST-COS quasar 
spectra \citep{Peeples17}, coupled with luminous red galaxies from the Sloan Digital Sky Survey \citep[SDSS;][]{York_SDSS}.
This search 
yielded a dataset comprising 94 background quasars probing approximately 3193 foreground luminous red
galaxies (LRGs) up to an impact parameter of $\sim 10$ pMpc. To gauge the potential of this 
dataset, we generated a simplified mock representation of these quasar-LRG pairs based on 
simulations. Our analysis reveals that if the actual feedback mechanisms resemble those presented 
in the IllustrisTNG simulation, then the observed $\rm dN/dz$ ($r_{\perp}$) values derived from 
these 3193 quasar-LRG pairs can effectively distinguish this scenario from the Illustris feedback 
model. The statistical significance of this distinction exceeds 12 $\sigma$. This outcome 
underscores the potential of $\rm dN/dz$ ($r_{\perp}$) as a robust tool for constraining AGN 
feedback models. In our forthcoming work, we intend to take a step further by measuring 
$\rm dN/dz$ ($r_{\perp}$) from archival HST-COS data. This effort will shed light on 
which feedback models align with observed $\rm dN/dz$ ($r_{\perp}$), providing further insights into the 
complex interplay between AGN feedback and galaxy formation.

The paper is organized as follows. In Section~\ref{sec.all_sims}, we provide an overview of the simulations used 
in this study. Section~\ref{sec.method} outlines our methodology, including the selection of halos hosting 
LRGs, the generation of realistic \lya~forest spectra probing these galaxies, 
and our Voigt-profile fitting procedure. In Section~\ref{sec.results}, we present the results of our analysis, 
encompassing various statistical measures, and conduct a feasibility analysis regarding the use 
of $\rm dN/dz$ ($r_{\perp}$) statistics on archival HST-COS data. Finally, in Section~\ref{sec.summary} we 
summarise the results of the paper.

\section{Simulations}\label{sec.all_sims}
Following \citetalias{Khaire23}, we use Illustris \citep{Vogelsberger14, Genel14} and 
the IllustrisTNG \citep{IllustrisTNG,Pillepich18} galaxy formation simulations to
study the impact of AGN feedback on the gas around halos of galaxies. 
Because both simulations are run with the same initial condition, we 
identify the same dark matter halos in both to investigate the difference in gas distribution surrounding them. 
Both simulations are also run with the same underlying {\sc arepo} code \citep{Springel10}, but as compared to Illustris the ideal magnetohydrodynamic calculations were used for IllustrisTNG. These simulations incorporate an array of astrophysical processes needed for galaxy formation and modeling the \lya~forest. These include star formation \citet{TNG_Stars}, 
feedback from both stars and AGN, the influence of galactic winds \citep[]{TNG_outflows}, the process of 
chemical enrichment \citep{TNG_metals}, photoionization heating using the UV background model of \citet{FG09}, and 
various cooling mechanisms including metal cooling.

The main difference between Illustris and IllustrisTNG simulation is the feedback 
prescription. Illustris failed to reproduce a few critical observations such as 
the distribution of red and blue galaxies \citep{TNG_galaxy_color}, 
populations of discs among galaxies \citep[]{TNG_Gas_disc}, and 
the gas mass fraction within $R_{500c}$ of the halos. In order to alleviate these issues 
many of the feedback prescriptions were changed in the updated simulation; 
IllustrisTNG. For an in-depth account of these updates, we recommend a 
comprehensive summary outlined in Table 1 of \citet{Pillepich18}. 

The primary distinction in feedback prescription, which significantly impacts the gas 
surrounding massive galaxies and the IGM, lies in the implementation 
of radio mode AGN feedback. In the Illustris simulation, a bubble mode feedback approach was 
employed, where the energy released by central AGN was accumulated over time and then 
explosively released into the surrounding environment. This process generated large, hot gas 
bubbles around massive halos in Illustris and substantially reduced the gas content within 
the region of R$_{500c}$ surrounding these halos. Conversely, IllustrisTNG utilized a kinetic 
radio mode feedback mechanism, wherein the energy released from central AGNs was used to 
drive the winds by imparting stochastic momentum kicks to nearby particles. These distinct 
approaches to AGN feedback are anticipated to influence the gas properties around massive 
galaxies, prompting our investigation into whether these properties can effectively serve as 
probes for the feedback mechanisms themselves.

Apart from the distinct AGN feedback prescriptions, there exists a slight variation in the 
cosmological parameters applied in both simulations, as outlined in Table~\ref{tab.cosmo}. 
Although we anticipate that these minor differences in cosmological parameters would have an 
insignificant impact on the properties of gas within the IGM and CGM, 
we maintained the respective cosmological parameters in our 
analysis of the \lya~forest. This choice ensures a fair and consistent comparison 
between the two simulations.

%
\begin{table}
\centering
\caption{Cosmology parameters used in simulations }
\begin{tabular}{cccc}
\hline
Parameters & Illustris & IllustrisTNG  \\
\hline
${\Omega_m}$       &  0.2726 &  0.3089  \\
$\Omega_{\Lambda}$ &  0.7274 &  0.6911  \\
$\Omega_{b}$       &  0.0456 &  0.0486  \\ 
$h$                &  0.704  &  0.6774  \\
$\sigma_{8}$       &  0.809  &  0.8159  \\
$n_s$              &  0.963  &   0.97   \\
\end{tabular}
\label{tab.cosmo}
\end{table}

It is expected to have a large impact of feedback at lower redshifts, therefore, as in 
\citetalias{Khaire23}, we choose a simulation snapshot at $z=0.1$. We used
publically available snapshots at $z=0.1$ of both Illustris \footnote{Illustris:
\href{https://www.illustris-project.org}{https://www.illustris-project.org}}
and IllustrisTNG\footnote{IllustrisTNG:
\href{https://www.tng-project.org}{https://www.tng-project.org}}
having a box size of $75$ cMpc/h and 1820$^3$ dark matter and baryon particles each.
In order to study the gas around galaxies and then 
generate synthetic \lya~forest, we first convert the simulation snapshots of both 
Illustris and IllustrisTNG into 3D Cartesian grids. This is achieved by
depositing smoothed quantities such
as density, temperature, and velocities from Voronoi mesh output of simulations
onto  1820$^3$  grids. For calculating smoothed quantities we follow the standard
approach of using a Gaussian kernel with a radius equal to $2.5$ times the radius of
each Voronoi cell by assuming each Voronoi cell is spherical. 
As mentioned in \citetalias{Khaire23}, this way of fixing Gaussian kernel size 
and the choice of factor $2.5$ is arbitrary but often used for 
Illustris and IllustrisTNG simulations (private communication with D. Nelson).

\begin{figure*}
    \centering
    \begin{minipage}{0.495\textwidth}
        \centering
        \includegraphics[width=1.16\textwidth,keepaspectratio]{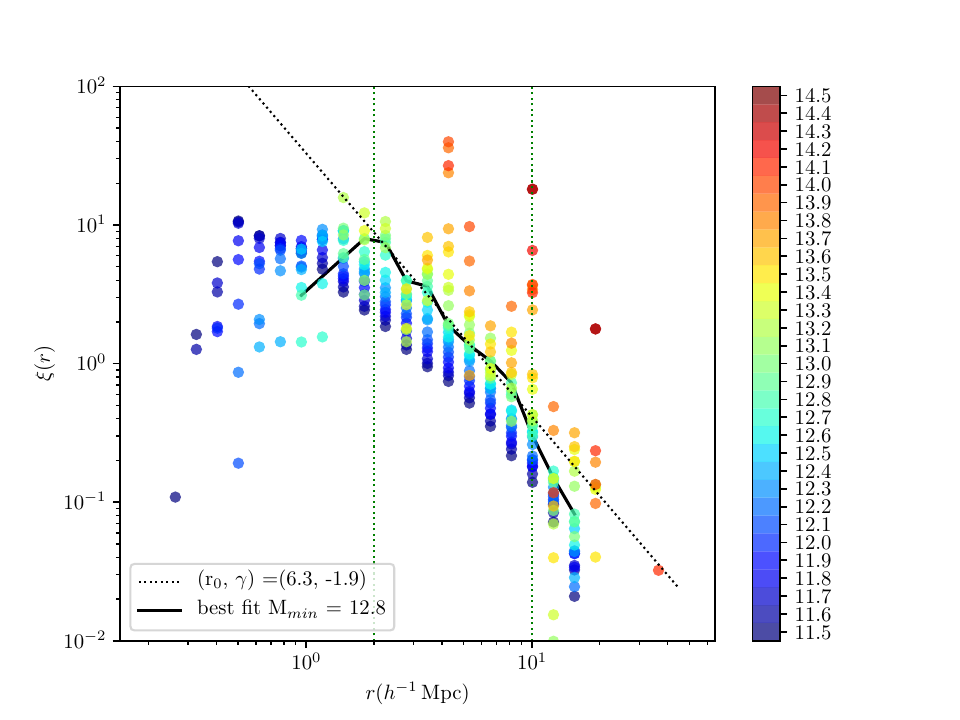} 
    \end{minipage}\hfill
    \begin{minipage}{0.495\textwidth}
        \centering
        \includegraphics[width=1.16\textwidth,keepaspectratio]{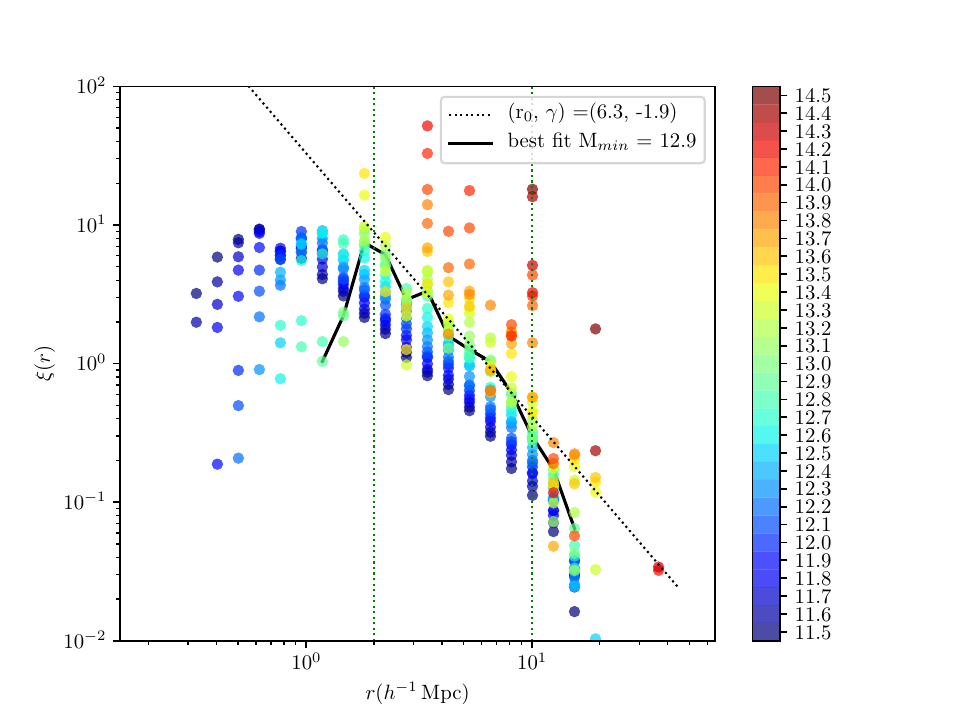} 
    \end{minipage}
    \caption{The projected 2-point correlation function $\xi(r)$ for Illustris (left-hand panel) 
    and IllustrisTNG halos (right-hand panel) with mass ($M > M_{\rm min}$), as indicated by color-bars. 
    The parametric fit form \citet{Guo15} measurements (\emph{dotted line with slope}) 
    match with the best fit (solid lines) $M_{\rm min} = 10^{12.8}$ $M_{\odot}$ for Illustris and $M_{\rm min} = 10^{12.9}$ $M_{\odot}$ 
    for IllustrisTNG. Vertical green dotted lines show the range of $r$ used for obtaining the best fit $M_{\rm min}$ values.}
\end{figure*}\label{fig.corr}

\subsection{Gas Around the Same Halo: An illustration}
Given the distinct radio-mode AGN feedback implementations in Illustris and 
IllustrisTNG, these simulations offer a valuable opportunity to explore the impact of 
AGN feedback on gas distribution around galaxy halos and within the IGM. 
Notably, since these simulations share identical initial conditions, 
enabling us to find the same halos for a direct comparison 
of differences in the distribution of gas around them. This is illustrated in 
Fig. \ref{fig.boxplot}. The figure shows a 2D slice of the hydrogen density 
and temperature profile around the same halo in both simulations. The slice has
a thickness equal to one grid cell (i.e of size $\sim 60$ ckpc).
The halos were identified by matching the coordinates of halos in the halo 
catalogs provided by Illustris and IllustrisTNG.  
The dashed circles at the center of each panel have a radius 
equal to the virial radius of the halo. The density and temperature profiles shown in
this example demonstrate that Illustris expels hot gas further from the central 
halo as compared to the IllustrisTNG (see density profile in left-hand panels), 
and also generates a huge $>10$ pMpc size shock-heated hot region identified by the transition from 
shock heated $\sim 10^6$ K gas to$10^4$ K  photoionized gas (see right-hand panels). The size of the bubble in 
Illustris halo is up to the size of 20 virial radii ($R_{\rm vir} = 0.57 $ pMpc 
for $10^{13.1}$ M$_{\odot}$ halo at $z=0.1$). This expulsion of gas also leads 
to small differences in the masses of halos that are calculated by summing the 
mass of all particles and cells within $R_{200c}$, resulting 
in slightly less mass for the Illustris halo ($10^{13.1}$ M$_{\odot}$) as compared 
to IllustrisTNG ($10^{13.2}$ M$_{\odot}$ ). 
Fig. \ref{fig.boxplot} shows that there are significant differences in 
the distribution of gas density and temperature around the same halos in both 
simulations. Only in some regions far away from halos, the density and 
temperature profiles are similar in both simulations (for e.g notice a filament at the bottom 
left corner of Fig.~\ref{fig.boxplot}). 

The differences in gas temperature and density surrounding halos provide a compelling reason 
to explore statistical metrics that can effectively differentiate between the feedback 
disparities in both simulations. The subsequent section outlines our methodologies for 
examining the gas properties around massive halos within these simulations. 

\section{Methods to generate synthetic Lyman alpha forest around massive halos}\label{sec.method}
In this section, we describe the methodology employed for the selection of massive halos in 
both simulations, the generation of synthetic \lya~forest spectra in the IGM and in the 
vicinity of halos, the forward modeling approach applied to construct mock HST COS spectra, 
the automated Voigt profile fitting procedure, and the UV background calibration for both 
simulations.

\subsection{Halo mass selection for LRG hosts}\label{appendixA}

In this study, our primary objective is to assess how AGN feedback influences the gas 
surrounding halos within the Illustris and IllustrisTNG simulations. Given that AGN feedback 
is expected to suppress star formation in massive galaxies, particularly LRGs, we focus on 
massive halos capable of hosting LRGs. Observational evidence supports the idea that LRGs 
tend to reside in massive halos, as indicated by their clustering properties and 
abundance. To select suitable halos of LRGs from the simulations, 
we employ a simple model based on a step-function halo-occupation distribution. 
In this model, we assume that simulated halos above a 
specified mass threshold $M_{\rm min}$, are hosts to LRGs.
We determine $M_{\rm min}$ such that halos 
with $M > M_{\rm min}$ in our $z=0.1$ Illustris and IlutrisTNG simulation boxes
match with the measured projected 2-point correlation function (2PCF) of 
LRGs at median redshift $z\sim 0.1$ obtained by \citet{Guo15}. 

Because it is not straightforward to estimate the projected 2PCF in simulations, 
we relate the redshift-space projected 2PCF, $w_p(r_p)$, to the real-space correlation 
function $\xi (r_p)$ by following  \citet{Zehavi11},
\begin{equation*}
    w_p (r_p) = 2\int^{\infty}_{r_p} rdr\xi(r) \big(r^2 -r_p^2\big)^{1/2}.
\end{equation*}
Then, for a power law $\xi (r_p) = (r/r_0)^{-\gamma}$ one obtains
\begin{equation*}
    w_p (r_p) = r_p \Big(\frac{r_p}{r_0} \Big)^{-\gamma} \Gamma(1/2) \Gamma\Big(\frac{\gamma-1}{2}\Big)/\Gamma(\gamma/2).
\end{equation*}
Using these relations we fit the $w_p(r_p)$ of \citet{Guo15} for galaxies
with magnitudes in SDSS $r$ band $M_r < -21$. Our simple chi-by-eye fit gives 
us parameters $r_0 =6.3$ and $\gamma = 1.9$. These values are also consistent
with the fit obtained by \citet[][$r_0 =6.46$ and $\gamma = 1.9$ when using only 
diagonal elements for fit]{Zehavi11} for the galaxies with same $M_r$ conditions.  

We proceed by selecting halos with a mass $M > M_{\rm th}$ from both simulations as given in their respective catalogs.
Subsequently, we compute the real-space correlation function, $\xi(r)$, 
for these halos using equispaced logarithmic bins over the range 
$0.01 < r \, (h^{-1} \, \text{Mpc}) < 40$. The resulting correlations are 
shown in Fig.~\ref{fig.corr}, where the color bar represents $\xi(r)$ values 
corresponding to different $M_{\rm th}$. 
We compare this $\xi (r)$  with our power-law fit to the $\xi {r}$ from 
\citet{Guo15} projected 2PCF measurements (dotted line in Fig.~\ref{fig.corr}) and
estimate the $M_{\rm min} = M_{\rm th}$ for which the chi-square value
is minimum. For minimizing chi-square we use bins of  
$0.01 <r\, (h^{-1} \, {\rm Mpc})< 40$.
This analysis yields values of $M_{\rm min} = 10^{12.8}$ $M_{\odot}$ for Illustris 
and $M_{\rm min} = 10^{12.9}$ $M_{\odot}$ for IllustrisTNG. 
These $M_{\rm min}$ values align closely with the $M_{\rm min} = 10^{12.78 \pm 0.11}$ 
$M_{\odot}$ value derived by \citet{Guo15} for $M_r < -21$, employing more 
sophisticated halo occupation distribution modeling (refer to their table 2). 
These $M_{\rm min}$ values yield a total of 296 halos in Illustris 
(for $M > 10^{12.8}$ $M_{\odot}$) and 305 halos in 
IllustrisTNG (for $M > 10^{12.9}$ $M_{\odot}$) at $z=0.1$.

\subsection{Generating Synthetic \lya~Forest around the halos and in the IGM}\label{sec.moc}
For generating synthetic \lya~forest, we need the temperature, density, and velocity of
cells along the sightlines and the H~{\sc i} photoionization rate ($\Gamma_{\rm HI}$)
from the UV background to determine the neutral fraction of hydrogen. 
We generate two types of sightlines: one set passes in proximity to our chosen halos that 
host LRGs, known as halo sightlines, while the other set probes the IGM and is referred to as 
IGM sightlines. The IGM sightlines are just sightlines generated at random being agnostic 
to positions of halos. 

For the IGM sightlines, we adopt the procedure outlined in \citetalias{Khaire23}.
We draw $10^5$ sightlines from one side of the simulation 
box to the other, parallel to an arbitrarily chosen $z-$axis. 
The starting points of these sightlines are chosen randomly on the $x-y$ plane. 

To create halo sightlines, we generate sightlines around halos hosting LRGs, 
aligned parallel to a chosen $z$-axis, at various impact parameters 
($r_{\perp}$), representing the perpendicular distance from the halo center to the 
sightline. We establish 30 bins for impact parameters ranging from 0 to 90 pMpc, each with 
differing bin sizes. These bin sizes are designed to approximate equal logarithmic intervals 
in impact parameter space. When generating these sightlines, we ensure that the maximum 
number of sightlines within an impact parameter bin around all LRG halos totals $10^5$.
To achieve this, we calculate the number of sightlines to be drawn around each halo in each 
impact parameter bin, which is given by $10^5/N_h$, where $N_h$ represents the total number 
of halos in each simulation box with masses greater than $M_{\rm min}$. Subsequently, for 
each impact parameter bin surrounding each halo, we randomly generate $10^5/N_h$ sightlines 
that traverse the available cells within the $x-y$ plane.
However, for the impact parameter bins closest to the halos, this value of $10^5/N_h$ is 
smaller than the number of available cells in the $x-y$ plane. 
Only in such cases, we generate sightlines that pass through all available cells within the 
annular ring defined by the edges of the respective impact parameter bins. For instance, in 
smaller impact parameter bins at $r_{\perp} < 0.5$ cMpc with a width of 
approximately 100 ckpc, the number of sightlines is less than $10^5$, and thus these 
sightlines are correlated. After creating the sightlines, we use periodic boundary 
conditions to center them at the real-space coordinate positions of the halos.

Along these halo and IGM sightlines, we store the values of 
temperatures $T$, overdensity $\Delta$ and the velocity component parallel to 
line-of-sight ($v_z$). 
In addition to these quantities, we also need neutral hydrogen fractions 
to generate the synthetic \lya~forest spectra.
For that, we solve for a neutral fraction in ionization equilibrium 
including both collisional ionization and photoionization. At every cell
along a line-of-sight, we use $T$, $\Delta$, 
and $\Gamma_{\rm HI}$ to calculate the neutral fractions.
We assume a constant value of $\Gamma_{\rm HI}$ for the whole simulation box.
Specific values of  $\Gamma_{\rm HI}$ used in Illustris and IllustrisTNG are discussed 
in Section~\ref{sec.gamma}. For ionization equilibrium calculation we need an estimate of the electron density
which can have a contribution of up to $\sim 16$\% more from the ionization of helium gas. 
Following \citetalias{Khaire23}, we assume helium is predominantly in He~{\sc iii} state
and calculate the electron density. This approximation of complete ionization of helium
is justified given that we are working at $z=0.1$ which is almost 10 billion years after
the epoch of helium reionization at
$z\sim 3$ \citep{Shull10, Worseck11, Khaire17sed}. This assumption further helps in removing
two free parameters, He~{\sc i} and  He~{\sc ii} photoionization rates, from ionization 
equilibrium calculation.  Our calculations for ionization equilibrium incorporate revised 
cross-sections and recombination rates from \citet{Lukic15}. Additionally, we adopt 
the self-shielding approach outlined in \citet{Rahmati13} for dense cells.

After determining the neutral fraction of hydrogen along a sightline, 
following procedure in \citetalias{Khaire23} \citep[see alss][]{Teng22, Hu23_whim} we use the values of $T$ and $v_z$ at each cell and calculate the \lya~optical depth $\tau$ at each cell
by adding all the real space contributions to the redshift-based \lya~optical depth using complete Voight profile \citep[with approximation provided in][]{Tepper06} arising 
from real-space cells. The continuum normalized flux is then just $F=e^{-\tau}$ along each sightline. We call these sightlines perfect \lya~forest sightlines. 
Pixel separation and hence the resolution of these sightlines is $4.12$ km s$^{-1}$ arising because we grid each simulation box into $1820^3$ cells.
This resolution of perfect sightlines is $\sim 4$ times finer 
than the resolution of HST-COS data ($\sim 15-20$ km s$^{-1}$).

\begin{figure*}
\includegraphics[width=0.98\textwidth,keepaspectratio]{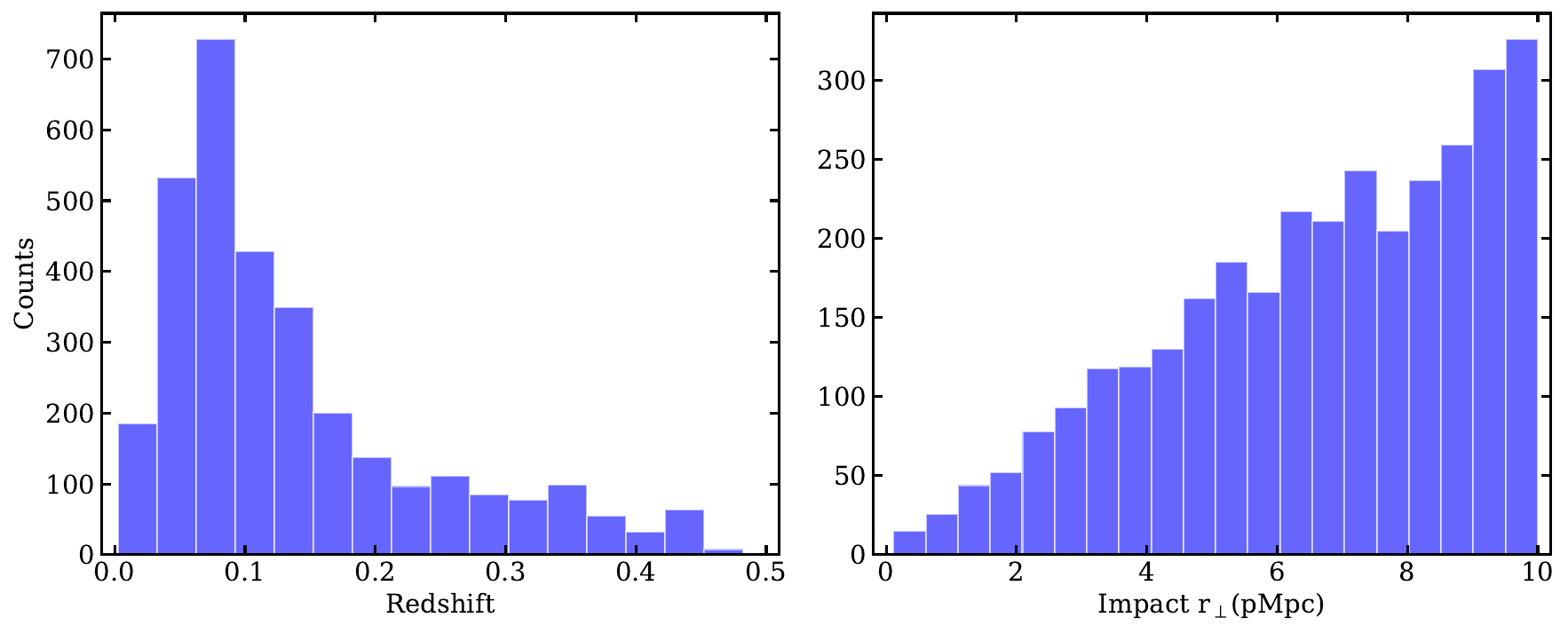}
\caption{Redshift (left) and impact parameter 
  ($r_\perp$, right) distribution of the 3193 foreground SDSS LRGs as probed by 94 background quasar spectra
  from the HSLA. These sightlines are selected to have spectra with S/N$>5$ per COS 
  resolution element in the \lya~forest region 
  (with sample mean of S/N $\,= 12.8$ per COS resolution element).}
\label{fig.data_archival}
\end{figure*}
%
\subsection{Forward models to determine photoionization rates}\label{sec.gamma}
In \citetalias{Khaire23}, we emphasized the significance of adjusting $\Gamma_{\rm HI}$ 
in simulations to facilitate comparisons with low-z IGM observations 
\citepalias[refer to section 4 of][]{Khaire23}. This adjustment becomes especially crucial 
when comparing two simulations that exhibit distinct fractions of diffuse, low-temperature 
($T< 10^5$ K) gas responsible for observed \lya~absorption. As outlined in 
\citetalias{Khaire23}, Illustris contains 23.2\% of baryonic mass in this diffuse \lya~phase, 
whereas IllustrisTNG has 38.5\%. Therefore, we applied the same methodology detailed in 
\citetalias{Khaire23} to determine $\Gamma_{\rm HI}$ for both Illustris and IllustrisTNG. 
A concise summary of this procedure is provided below.

We created forward models with IGM sightlines drawn from simulations following the 
noise properties and convolution of lifetime position-dependent HST-COS LSF for  
\citet{Danforth14} dataset in redshift bin $0.06 < z < 0.16$. This redshift bin contains 32 
quasar spectra with S/N $> 10$ per pixel (of size $6$ km s$^{-1}$) 
and the median of \lya~forest region covered in the redshift bin falls at $z=0.1$. 
While creating forward models we stitched our IGM sightlines to cover the 
\lya~redshift path of each observed quasar spectra.   
We generate these forward models for different values of $\Gamma_{\rm HI}$. 
We conducted Voigt profile fitting to 
all the \lya~lines falling in this redshift bin in the data as well as a set of 
forward models with our automatic python wrapper \citep[developed in][]{Hiss18} that runs 
VPfit {v10.2}
\citep{Carswell14}\footnote{see 
\href{http://www.ast.cam.ac.uk/~rfc/vpfit.html}{http://www.ast.cam.ac.uk/~rfc/vpfit.html}}. 
Using the results of fitting we calculate the \lya~line number density $dN/dz$, 
i.e. the number of \lya~lines within a column density range per unit redshift for 
both simulations and the data. To be consistent with the results of \citetalias{Khaire23}, 
we choose the column density range for $dN/dz$ to be 
$12< {\rm log} N_{\rm H I} ({\rm cm ^{-2}})<14.5$. 
In this specific column density range, we obtained $dN/dz = 205$ for the \citet{Danforth14} 
dataset within the redshift bin $0.06 < z < 0.16$. 
We determine the best fit $\Gamma_{\rm HI}$  for both simulations that matched well 
with the $dN/dz$ of the data. 
We found  $\Gamma_{\rm HI} = 3.1 \times 10^{-12}$ s$^{-1}$ for Illustris and  
$ 7.73 \times 10^{-12}$ s$^{-1}$ for IllustirsTNG, 
as found in  (refer to Section 4 of) \citetalias{Khaire23}. 
For the subsequent analysis presented in this paper, 
we use these values of $\Gamma_{\rm HI}$ for 
both IGM and halo sightlines, unless otherwise specified.

\subsection{Forward models for halo and IGM sightlines motivated by archival data}\label{sec.fwd_halos}

Motivated by large differences in the temperature and density of gas around halos in the 
Illustris and IllustrisTNG simulation, we wish to investigate if \lya~forest around those 
halos shows significant differences so that we can use it to constrain feedback models. 
Moreover, in order to study the feasibility of using \lya~forest around halos to constrain 
the feedback, we want to find out the properties of the data that is already available in the 
archive so that we can approximate our forward models accordingly. 
Therefore, we mined the HST spectroscopic legacy archive \citep[HSLA;][]{Peeples17} and 
searched for background quasar spectra which have  
\lya~forest with ${\rm S\slash N} > 5$ per resolution element of
COS (with FWHM of $\sim 18$ km s$^{-1}$ ). Then we used the SDSS DR12 LRG catalog
\citep{Tojeiro12} to cross-match these quasar sightlines
with LRGs. We identified 3193 quasar-LRG pairs where each 
background quasar probes the Ly$\alpha$ forest at the redshift of one or more foreground LRGs at impact parameters in the range 
$0.1 \,{\rm pMpc}$$\,< r_{\perp} <\,$$10\,{\rm pMpc}$.
This means we select 
LRGs with redshift $z_{LRG} < z_q$, where $z_q$ is the redshift of the background quasar. 
Our selected 3193 LRGs are being probed by 94 HSLA quasar spectra observed by FUV gratings of HST COS. The mean S/N per resolution element of the COS in the Ly$\alpha$ forest across the entire dataset is $12.8$ which corresponds to S/N of 7.4 per pixel of size $\Delta v = 6$ km s$^{-1}$ 
as used in \citet{Danforth14} binned data.
The redshift distribution of these foreground 3193 galaxies and their impact parameter from the 94 background quasar sightline probing \lya~forest is plotted in Fig.~\ref{fig.data_archival}. The redshift distribution of galaxies peaks at $z\sim 0.1$ and therefore aligns well with our simulation box redshift. 
Our choice of impact parameter range up to $10$ pMpc is motivated by the significant differences seen in the temperature and density profile of massive halos up to 10-20 pMpc as shown in Fig.~\ref{fig.boxplot}.

Our selected quasar-LRG sample includes data from past HST COS programs that specifically targeted quasars near LRGs
\citep{Chen18, Chen19, Zahedy19, Berg19}. 
However, it is worth noting that these prior studies primarily focused on regions within $r_{\perp} \lesssim R_{\rm vir} \simeq 0.5~{\rm pMpc}$, concentrating on the CGM and metal lines. 
In contrast, our investigation extends to much larger impact parameters, reaching up to $10$ pMpc.

Given these properties of quasar-LRG pais from archival data, we adopt a simple approach of modeling \lya~forest in the halo and IGM sightlines with S/N $=7$ per pixel (of $\Delta v = 6$ km s$^{-1}$).
Furthermore, for simplicity, we assume all 3193 LRGs are at a single redshift $z=0.1$ which is also the median of the redshift distribution  
(see left-hand panel of Fig.~\ref{fig.data_archival}), 
so that we can just analyze a single snapshot of both
simulations used here at $z=0.1$. For the generation of these forward models, we first convolve our perfect sightlines with HST-COS G130M grating LSF that covers \lya~forest at $z=0.1$ 
and then rebin the spectrum to match the pixel separation 
given in \citet{Danforth14} paper (i.e with a pixel velocity scale of $6$ km s$^{-1}$) and then add Gaussian random noise at each pixel to get the S/N of $7$ per pixel. 
%
\begin{figure*}
\includegraphics[width=0.98\textwidth,keepaspectratio]{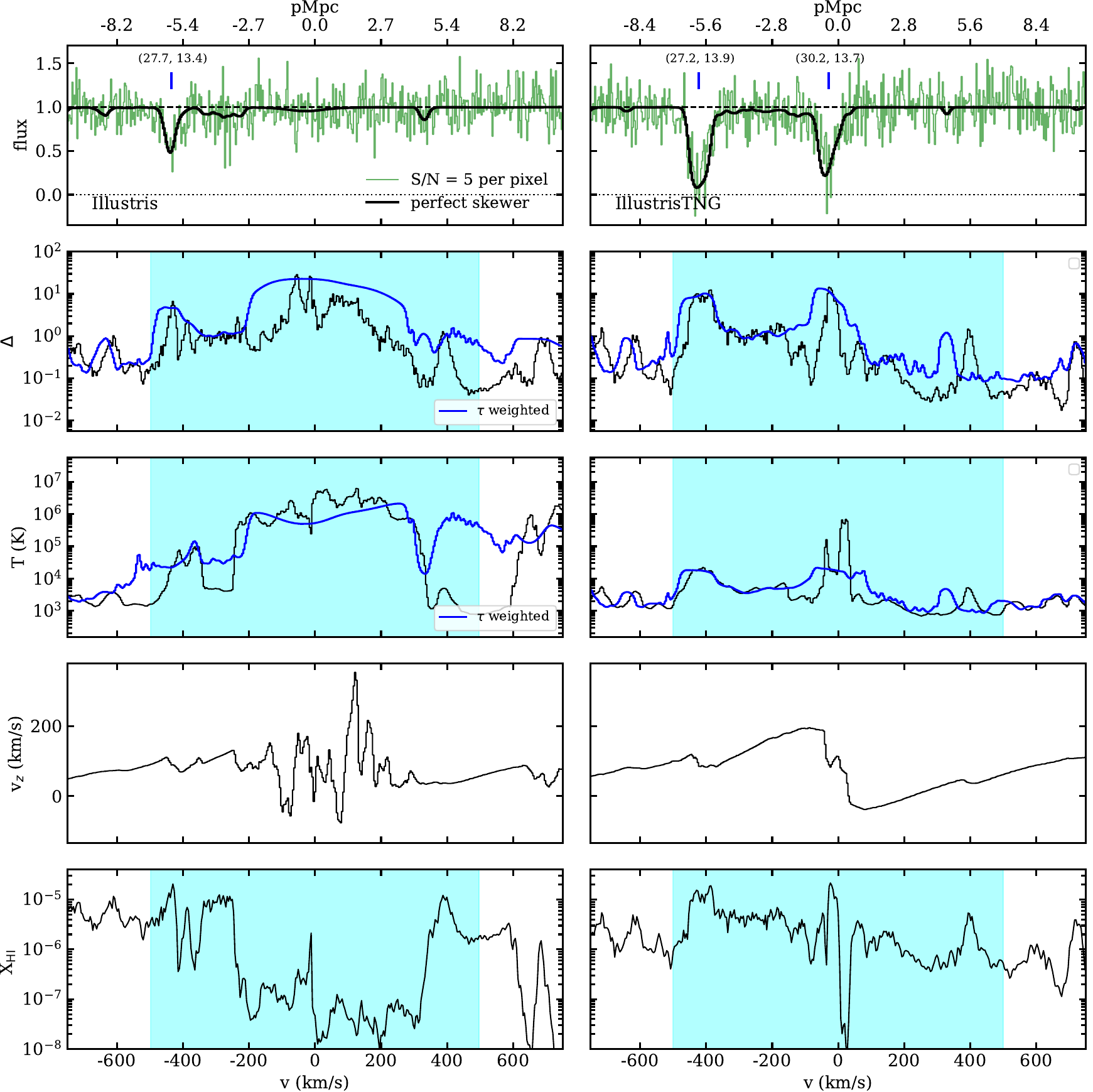}
\caption{
A representative example of \lya~forest sightlines at $ {r_{\perp} \sim 1.3}$ pMpc
around the same halo in Illustris (left-hand panels) and IllustrisTNG (right-hand
panels).
From the top, black lines show the normalized flux (perfect spectra), 
overdensity ($\Delta$), temperature, line of sight velocity ($v_z$) and 
\HI~fraction along the sightline. Green histograms on top panels show 
the HST-COS forward modeled sightlines with S/N $= 7$ per pixel 
(of size $\Delta v = 6$ km s$^{-1}$).
The blue ticks indicate VPfit 
identified \lya-absorption lines and legends in the bracket 
show the values of fitted $b$ (km s$^{-1}$) and log $(N_{\rm HI}/\text{cm}^{-2})$, 
respectively. 
Blue curves in the second and third panels from the top show optical depth
weighted overdensities and temperature respectively. 
Because the gas is much hotter for Illustris just one \lya~absorption line has been 
identified in the forward-modeled spectra as 
compared to two lines for IllustrisTNG within $\pm 500$
km s$^{-1}$ window highlighted by shaded cyan color. 
Different sizes of the sightlines on the left and
right panel (as noted on top $x-$axis) is a consequence of the different values
of Hubble constants used in the simulations (see Table~\ref{tab.cosmo}).
}
\label{fig.halo_sightlines}
\end{figure*}
%

After creating forward-modelled halo sightlines, we fit them 
with our automated VPfit code to obtain $b$ 
and \nhi~for all the \lya~absorption lines falling within 
$\pm 500$ km s$^{-1}$ centered at the position of halos. Our choice of this 
velocity width, $\pm 500$ km s$^{-1}$ 
along the sight-line, is an arbitrary choice only motivated to keep a large 
enough spectral chunk to probe \lya~forest liens far away from halos. 
Additionally, this choice allows us to account for lines that may be offset from the center of the halos due to the high-velocity gradients near these structures.
In order to choose  $\pm 500$ km s$^{-1}$ velocity windows, 
the center of the sightline is determined by the real space position of 
the halo on $z-$axis, subtracted by the $z-$component of the peculiar 
velocity of halo. This centering method resembles the procedure followed 
in the observations of background quasar absorption lines probing the halo of 
a foreground galaxy. 

In Fig.~\ref{fig.halo_sightlines}, we show a portion of
two such examples sightlines passing through the same locations around the 
same halos in Illustris (left-hand panel) and IllustrisTNG 
(right-hand panel) at an impact parameter $r_{\perp} = 1.3$ pMpc.
The top panel shows our forward model as well as perfect \lya~absorption 
spectra.  Blue ticks show the \lya~absorption
lines identified by our automated VPfit, while the two numbers in bracket indicate $b$ in km s$^{-1}$ and $\log (N_{\rm HI}/\text{cm}^{-2})$. 
Within a velocity window $\pm 500$ km s$^{-1}$ centered on the position of halos 
(cyan shaded region), VPfit identifies one \lya~absorption lines
for the sightline passing through the halo of Illustris and two
\lya~absorption lines for sightline passing through the halo of IllustrisTNG.
Fig.~\ref{fig.halo_sightlines}
also shows the temperature and density, and corresponding
optical depth weighted quantities (refer to equation 1 of \citetalias{Khaire23}), 
along the sightlines. 
It shows that the gas is much hotter ($10^6 - 10^7$ K)
around the Illustris halos and extends up to
a large distance because of the aggressive bubble-radio mode feedback as compared 
to the IllustrisTNG. This is responsible for the difference in the number 
of lines identified by VPfit. The velocity $v_z$ 
parallel to $z-$axis, as shown in the fourth panel (from the top) of the figure,
illustrates the regions that are affected by the 
feedback. Large velocity gradients in Illustris as compared to IllustrisTNG show
that the strong feedback in the former is displacing gas violently around the halo.
Whereas far away from the halo, at large $v_z$, velocity profiles of both simulations are similar.
The last panel in the figure indicates the ionization fraction of 
neutral hydrogen $X_{\rm HI}$ for each cell. 
The $X_{\rm HI}$ illustrates the fact that in the region around halos,
it is the combination of high over-densities and low temperature that give rise to
higher $X_{\rm HI}$ values, of the order of $10^{-5}$, 
responsible for imprinting the \lya~absorption lines on the sightlines. 
Given that the temperatures are lower around IllustrisTNG halos, more
\lya~absorption lines are imprinted on sightlines. 
There are also some very broad but weak lines appearing on the perfect sightline 
because of very high-density high-temperature gas near the halo, however,  they 
are not identified by VPfit in forward-modeled spectra because of the finite S/N.

As shown in the example (Fig.~\ref{fig.halo_sightlines}), we Voigt profile fit all 
the halo sightlines as well as IGM sightlines and obtain the $b$ and $N_{\rm HI}$ for 
each \lya~line detected by our automated routine. Using these we study different statistics 
of gas around halos as a function of $r_{\perp}$. These 
are discussed in the following section.

\section{Impact of AGN feedback on the gas around LRGs}\label{sec.results}
In this section, we describe the results of our Voigt profile fitting of the \lya~lines
near halos at different impact parameters and study the feasibility of using those as 
a diagnostic tool to probe the impact of AGN feedback. For this, we use the joint and 
marginalized distribution of $b$ and $N_{\rm HI}$ and $dN/dz$ with impact parameter 
$r_{\perp}$ as summary statistics. The results of those are described in the following subsections. 

\begin{figure*}
\includegraphics[width=0.98\textwidth,keepaspectratio]{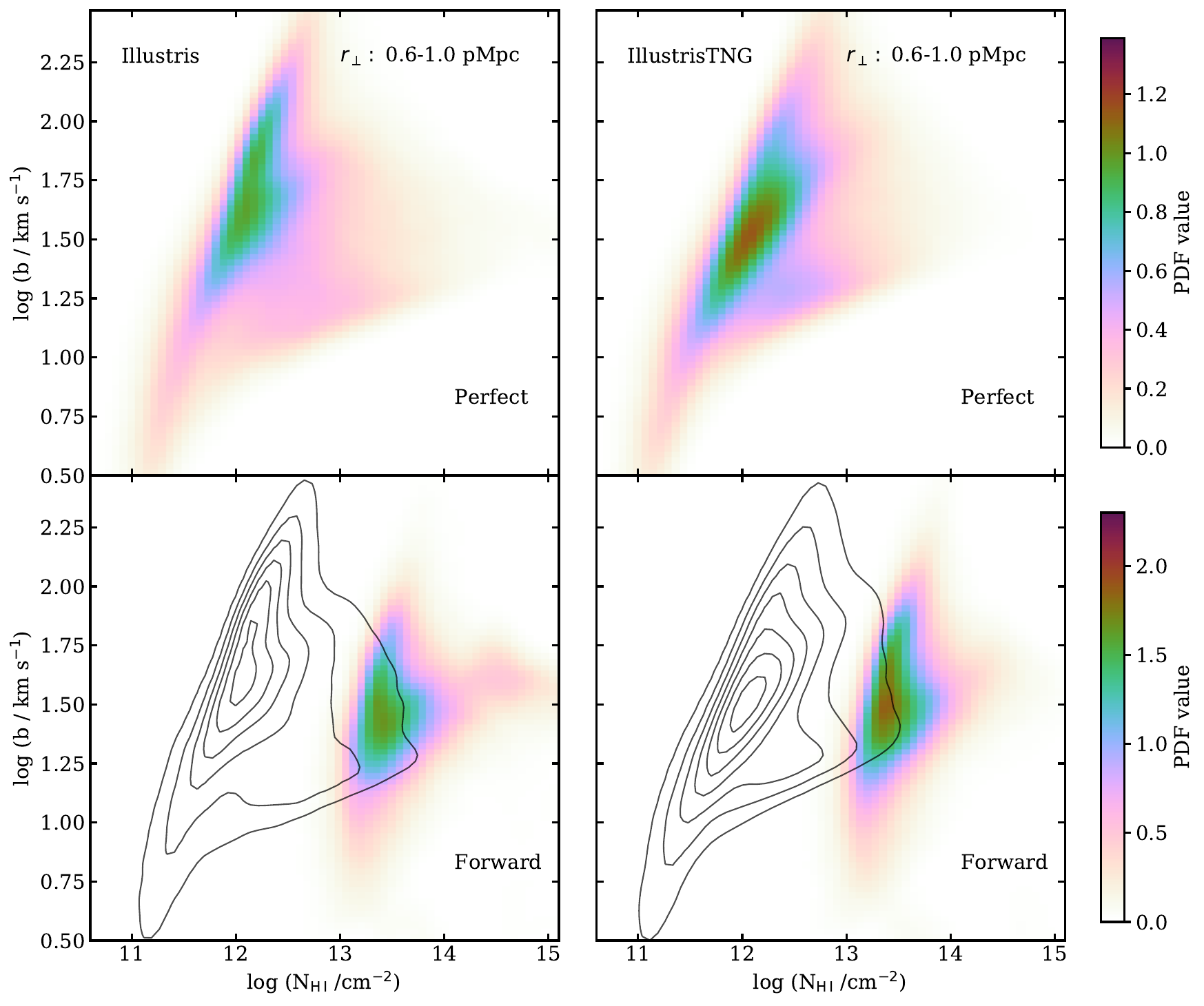}
\caption{The KDE estimated 2D $b-$\nhi~distribution for the halo sightlines  
(within $\pm 500$ km $s^{-1}$) drawn from  the impact parameter bin 
$0.6 <r{_\perp}< 1$ pMpc  
of Illustris (left-hand panel)
and the IllustrisTNG (right-hand panel) halos for perfect (top) 
and forward (bottom) models. 
The distributions from perfect and forward models look similar expect
the normalization shown with colors. 
In the bottom panel, we show $b-$\nhi~distribution from perfect models 
with contours. The forward models shift $b-$\nhi~distribution  towards higher 
\nhi~and lower $b$ values in both simulations. }
\label{fig.kde_small_impact}
\end{figure*}
\subsection{The $b-$\nhi~distribution around halos}\label{sec.bn_dist}
We select a central velocity window of 
$\Delta {v_{z}} = \pm 500$ km s$^{-1}$ of halo sightlines
and store the $b$ and \nhi~values of the \lya~absorption 
lines identified and fitted by VPfit for forward modeled as
well as perfect \lya~forest spectra. 
As in the \citetalias{Khaire23}, for perfect spectra, we add a 
small Gaussian random noise (which gives $\text{S/N} = 100$ per pixel of size $4.2$ km s$^{-1}$) so that VPfit can robustly identify lines. Without such a noise, VPfit fails to identify many \lya~lines.
This is because VPfit identifies lines bound by flux equal to unity 
and for most of the regions in the perfect sightlines the flux is smaller 
than unity by a tiny amount of 10$^{-3}$ to 10$^{-5}$. 
We mitigate this issue by adding a very small noise so that the flux 
becomes unity and VPfit can 
identify lines.  Once we fit those lines, we use the $b$ and \nhi~values to  
calculate 2D $b-$\nhi~distribution with Gaussian KDE at different impact parameters around 
massive halos of Illustris and IllustrisTNG.

\begin{figure*}
\includegraphics[width=0.98\textwidth,keepaspectratio]{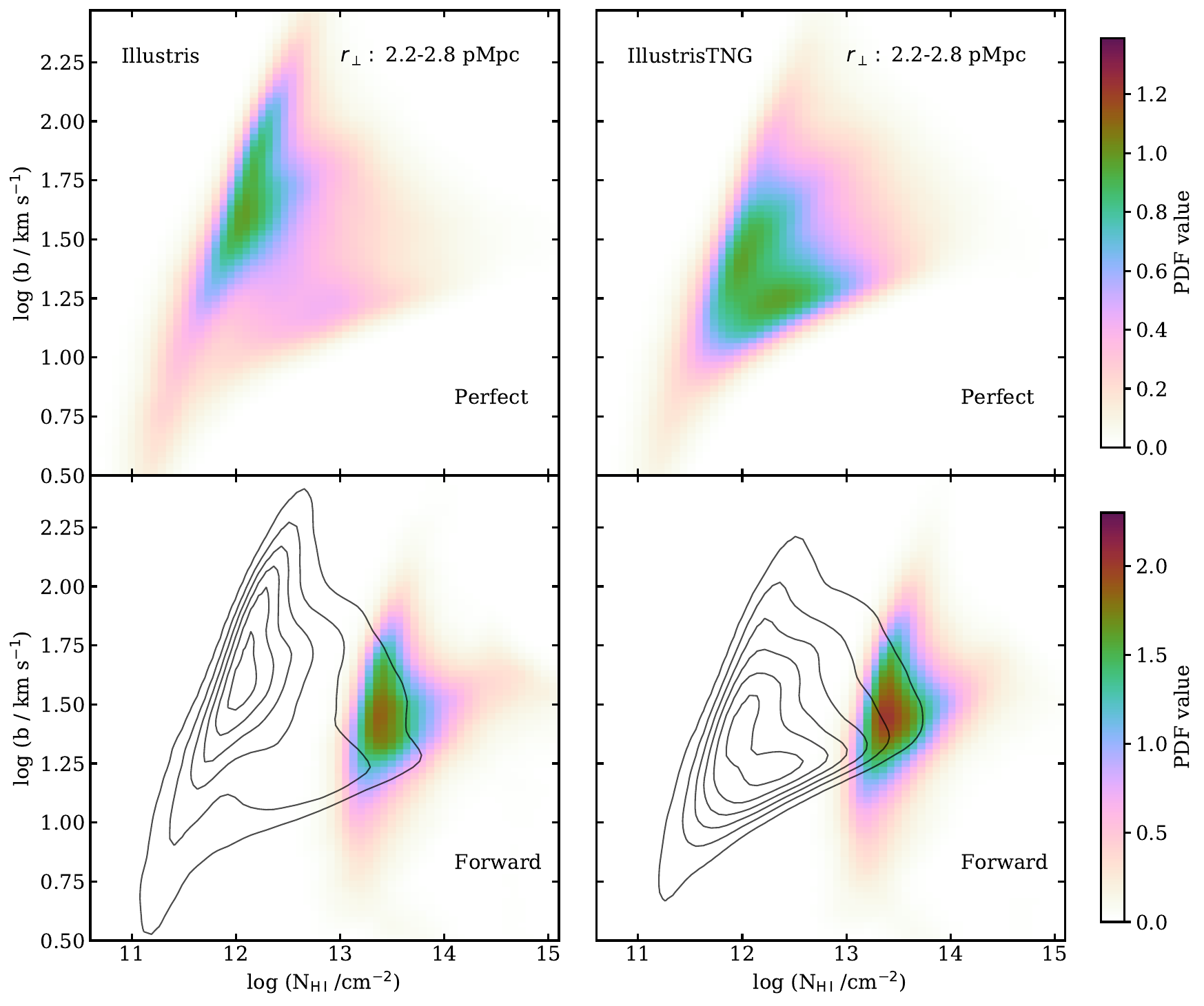}
\caption{
The KDE estimated 2D $b-$\nhi~distribution for the halo sightlines  
(within $\pm 500$ km $s^{-1}$) drawn from  the impact parameter bin 
$2.2 <r{_\perp}< 2.8$ pMpc  
of Illustris (left-hand panel)
and the IllustrisTNG (right-hand panel) halos for perfect (top) 
and forward (bottom) models. 
The effect of feedback can be seen
in the difference in the distributions (in top panels) for perfect sightlines. 
The hot gas in the Illustris
results in many absorption lines with low \nhi~high and $b$ values whereas
the  IllustrisTNG
shows both the low \nhi-high $b$ and high \nhi-low $b$ absorption lines. 
These differences in the shape of $b-$\nhi~get washed out in forward-modelled 
sightlines as shown in the bottom panel where the black contours indicate 
distribution from the perfect model shown on the top panel. However, the normalization of
distribution showing different numbers of lines,  
indicated by the color bars is still
significantly different even in the distribution from forward-modeled sightlines. }
\label{fig.kde_intermediate_impact}
\end{figure*}

\begin{figure*}
\includegraphics[width=0.98\textwidth,keepaspectratio]{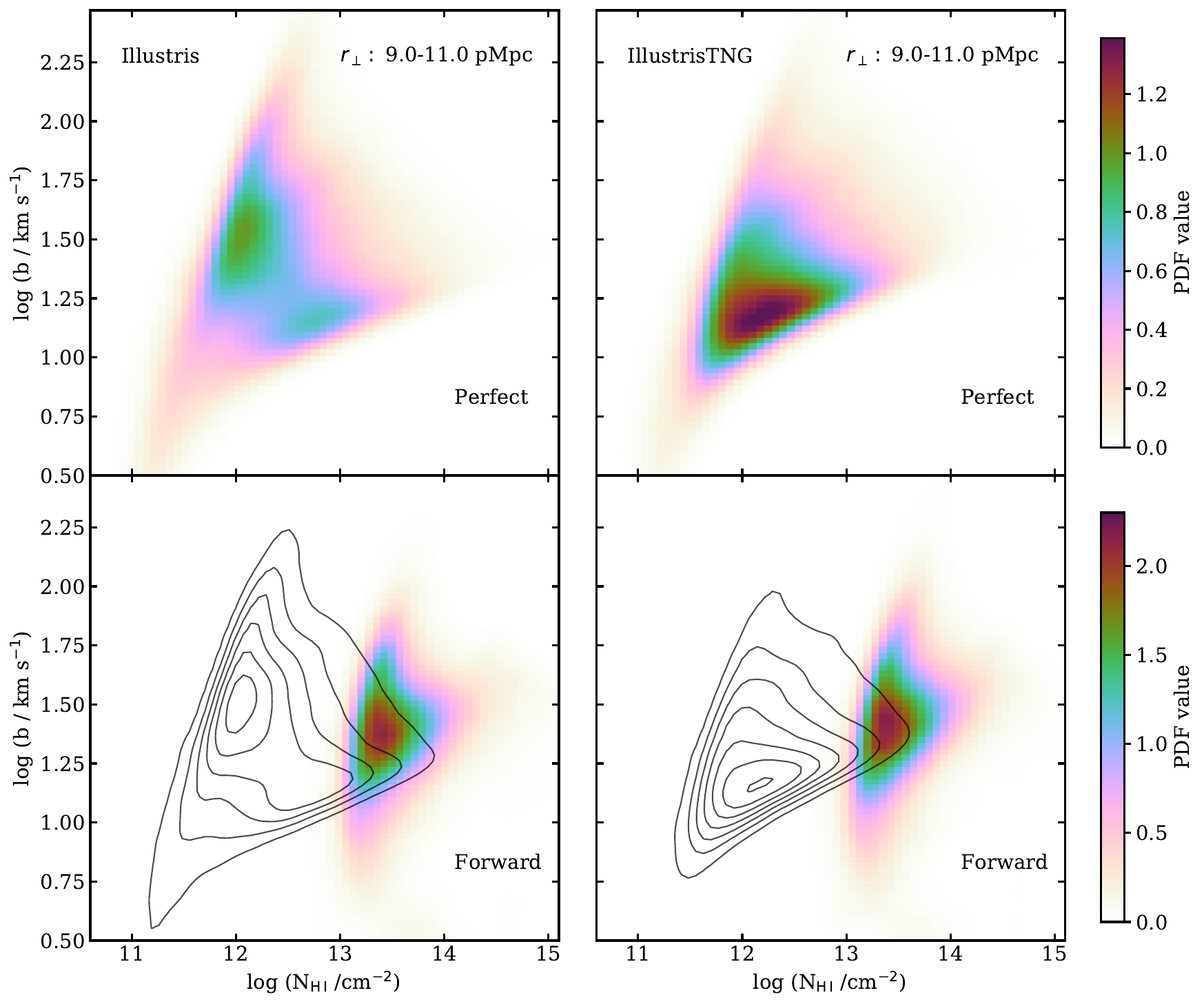}
\caption{The KDE estimated 2D $b-$\nhi~distribution for the halo sightlines  
(within $\pm 500$ km $s^{-1}$) drawn from  the impact parameter bin 
$9 <r{_\perp}< 11$ pMpc  
of Illustris (left-hand panel)
and the IllustrisTNG (right-hand panel) halos for perfect (top) 
and forward (bottom) models. 
The effect of feedback can be seen
in the difference in the distributions (in top panels) for perfect sightlines. 
As compared to the $b-$\nhi~distribution obtained for low 
(Fig.~\ref{fig.kde_small_impact}) and intermediate (Fig.~\ref{fig.kde_intermediate_impact})
$r{_\perp}$ bins there are more lines at high-\nhi~low lower-$b$ values in Illustris whereas
most of the lines in IllustrisTNG are at high-\nhi~low lower-$b$ values. 
For forward-modelled sightlines, the $b-$\nhi~distributions are indistinguishable and 
close to the distribution observed for the IGM (see \citetalias{Khaire23}). See Fig.~\ref{fig.bn_kde_apendix0} and~\ref{fig.bn_kde_apendix1} for $b-$\nhi~at other $r{_\perp}$ bins.}
\label{fig.kde_large_impact}
\end{figure*}

\begin{figure*}
\includegraphics[width=0.98\textwidth,keepaspectratio]{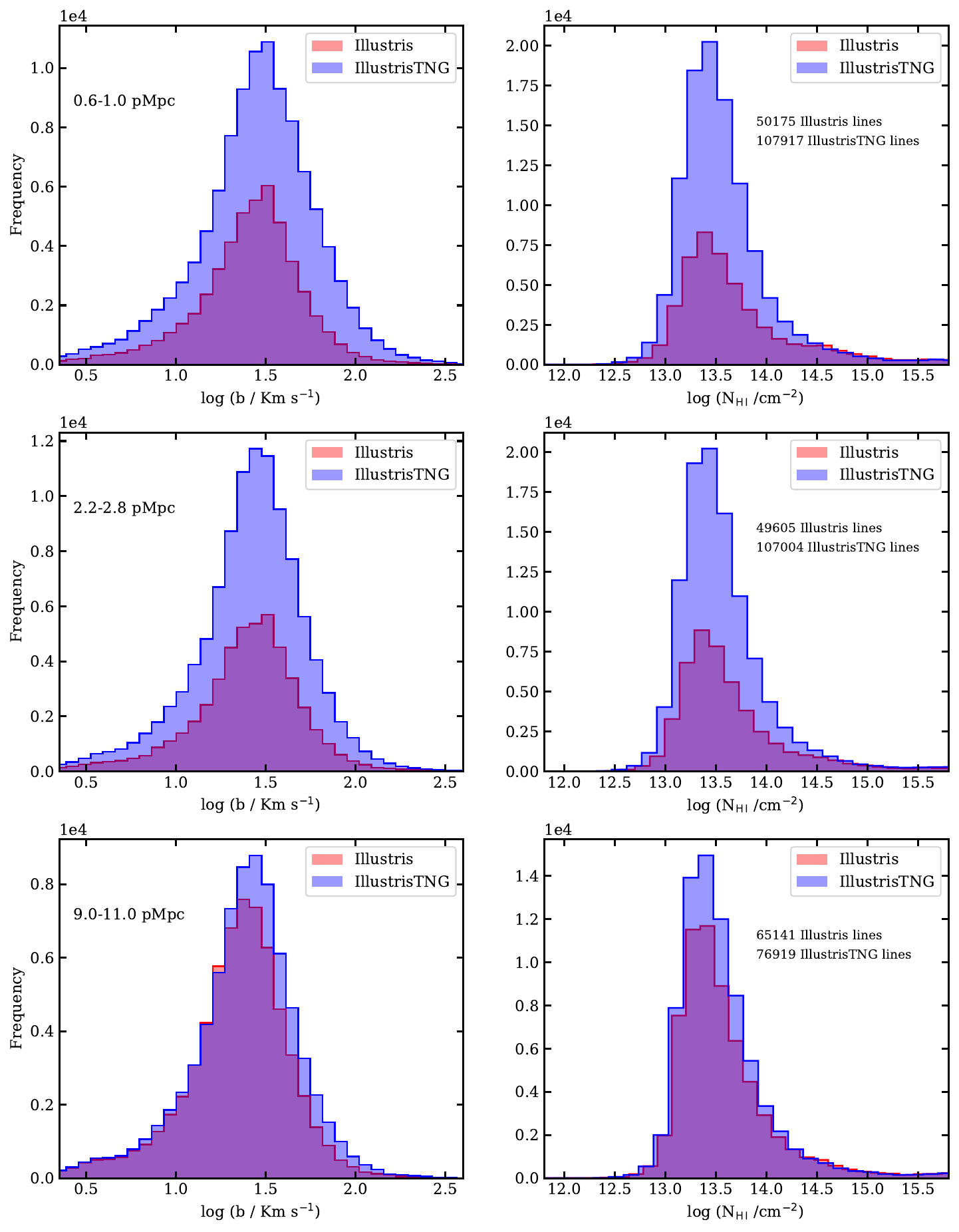}
\caption{The log $b$ (left) and log \nhi~(left) distribution 
from the forward modeled sightlines generated at a low impact parameter bin $0.6 <r_{\perp} < 1.0$ pMpc (top panel),  at an intermediate impact parameter bin $2.2 <r_{\perp} < 2.8$ pMpc (middle panel) and at high impact parameter bin $9 <r_{\perp} < 11$ pMpc (bottom panel). 
Distribution is not normalized and the number of total \lya~lines in each bin
are indicated in left-hand panels. Both simulations use the same number of sightlines ($\sim 10^5$)
in each $r_{\perp}$ bin. See Fig.~\ref{fig.halo_forward_hist_1} and \ref{fig.halo_forward_hist_2} for other $r_{\perp}$ bins. }
\label{fig.halo_forward_hist_0}
\end{figure*}

In Fig.~\ref{fig.kde_small_impact} we show the  normalized 
2D KDE $b-$\nhi~distributions for Illustris (left-hand panels) 
and IllustrisTNG (right-hand panels) at
an impact parameter bin of $0.6 <r{_\perp}< 1$ pMpc from selected halos. 
This impact bin is 1 to 2 Virial radii from the halos of the 
given $M_{\rm min}$ mass.
The top panels show the $b-$\nhi~distributions obtained with perfect sightlines 
(S/N $=100$ and  $\Delta v = 4.2$ km s$^{-1}$). The shape of the distribution is
qualitatively similar for both simulations showing a higher density of lines with 
$\log ( b / \text{km s}^{-1}) > 1.25$ and $\log (N_{\rm HI}/ \text{cm}^{-2}) > 11.5$. 
The similar shapes of the distribution suggest that at very close to the halos,
the temperature and density of gas in both simulations are similar 
(see for e.g Fig.~\ref{fig.boxplot}). However, the normalization of the
distribution which is indicated by the color bar in Fig.~\ref{fig.kde_large_impact}
shows that there are more lines in IllustrisTNG than in Illustris. 
Note that in these perfect spectra
VPfit could identify large $\log ( b / \text{km s}^{-1}$) upto $\sim 2.5$ (i.e $\sim 300$ km s$^{-1}$). 
In the bottom panel of Fig.~\ref{fig.kde_small_impact} 
we plot $b-$\nhi~distributions obtained with forward
modelled \lya~forest spectra. Here also, the shapes of distributions from both simulations
seem to agree qualitatively. For comparison with results from perfect 
sightlines, the bottom panels also show the $b-$\nhi~distributions from the perfect model
(as plotted in the top panels) with contours.  
As compared to those contours, the $b-$\nhi~distributions
from forward models shift towards higher values in $\log (N_{\rm HI}/ \text{cm}^{-2})  > 13$. Even here
with noisy forward models, the normalization is different between the two simulations 
showing that there are more \lya~lines in IllustrisTNG than in Illustris. 
However, in these noisy spectra
VPfit could identify $\log b$ values up to $\sim 2$ (i.e. $\sim 100$ km s${-1}$),
significantly smaller than in the case of perfect sightlines. This suggests that 
there are many broad lines that are too shallow to get detected by VPfit in 
forward-modelled spectra. Another interesting weak trend seen in these plots is 
there are more lines at higher $\log ( b / \text{km s}^{-1}) > 2$ values in 
IllustrisTNG than in Illustris. It is mainly
because the gas temperatures in IllustrisTNG especially for hot gas ($T>10^{5.5} k$)
are lower than Illustris  (as can be seen in Fig.~\ref{fig.boxplot} and temperature density 
phase diagram shown in \citetalias{Khaire23}) which can
imprint deeper broad lines that are easily detected by VPfit. 

Similar to  Fig.~\ref{fig.kde_small_impact}, 
in Fig.~\ref{fig.kde_intermediate_impact} we plot normalized 2D 
KDE $b-$\nhi~distributions at an intermediate impact parameter 
$2.2 <r{_\perp}< 2.8$ pMpc from massive halos in both simulations. 
Unlike the case with a small impact parameter (shown in Fig. ~\ref{fig.kde_small_impact}) 
the $b-$\nhi~distributions for perfect \lya~forest spectra show clear differences. 
IllustrisTNG $b-$\nhi~distribution is bimodal and it clearly shows two lobes
one at $\log ( b / \text{km s}^{-1}) >1.25$ and another at $\log ( b / \text{km s}^{-1}) < 1.25$, where former is mainly at lower column densities ($\log (N_{\rm HI}/ \text{cm}^{-2}) < 12.3$) and latter is at high column densities ($\log (N_{\rm HI}/ \text{cm}^{-2}) > 12$). Whereas Illustris just has one prominent region at  
$\log ( b / \text{km s}^{-1}) >1.25$ and it is remarkably similar to the one at lower impact parameter 
(see Fig.~\ref{fig.kde_small_impact}). 
This is mainly because there is no significant difference in the temperature 
and density profile in Illustris halos till such a large (up to $\sim 3$ pMpc)
impact parameters as illustrated in one example of temperature and density profile 
around halos in Fig.~\ref{fig.boxplot}. Note the highest $b$ values probed by 
IllustrisTNG is now lower than Illustris suggesting that the temperatures are
significantly lower around these impact parameters in IllustrisTNG. 
Bottom panels of Fig.~\ref{fig.kde_intermediate_impact} compare $b-$\nhi~distributions
from forward modeled \lya~forest spectra. The trends are similar to what is seen in 
the case of lower impact parameters (see, bottom panel of Fig.~\ref{fig.kde_small_impact}) but
there is a less prominent difference in the normalization.
It shows there is still a difference in the number of \lya~absorption lines but 
not as high as seen in the case of a lower-impact parameter.

Similar to  Fig.~\ref{fig.kde_small_impact} and Fig.~\ref{fig.kde_intermediate_impact}, 
in Fig.~\ref{fig.kde_large_impact} in we plot normalized 2D KDE $b-$\nhi~distributions
at a large impact parameter $9 <r{_\perp}< 11$ pMpc from massive halos in
both simulations. The $b-$\nhi~distributions for perfect \lya~forest spectra show 
clear differences between both simulations and they are in very good agreement with the 
ones obtained for IGM (as shown in Fig. 4 of \citetalias{Khaire23}). 
As compared to lower and intermediate impact parameters (Fig.~\ref{fig.kde_small_impact} 
and Fig.~\ref{fig.kde_intermediate_impact}) Illustris shows many lines with lower 
$b$ values ($\log ( b / \text{km s}^{-1}) < 1.75$) and there are many more lines with
high column densities $\log (N_{\rm HI}/ \text{cm}^{-2}) >12.5$ and small $b$ values ($\log ( b / \text{km s}^{-1}) < 1.25$).
In the case of IllustrisTNG however, the $b-$\nhi~distribution shows that most of the
\lya~absorption lines are at small $b$ values  $\log ( b / \text{km s}^{-1}) < 1.4$ and high column 
densities $\log (N_{\rm HI}/ \text{cm}^{-2}) >12$. As expected from density and temperature 
profiles of typical halos, a comparison of $b-$\nhi~distributions 
with impact parameters (Fig.~\ref{fig.kde_small_impact}, 
Fig.~\ref{fig.kde_intermediate_impact} and Fig.~\ref{fig.kde_large_impact}), 
shows that with increasing impact parameters the \lya~absorption lines become narrow  
(decrease in $b$ values) and contain more neutral hydrogen gas (increase in $N_{\rm HI}$). 
This trend is seen in both simulations but it is more prominent in the case of IllustrisTNG.  
To further illustrate such a trend we show $b-$\nhi~distributions at different impact 
parameters in Appendix (Fig.~\ref{fig.bn_kde_apendix0} and \ref{fig.bn_kde_apendix1}).   
Very close to the halos at lower impact parameters, the shape of the 
$b-$\nhi~distributions are similar in both simulations and they start differing
as the impact parameters are increased. The $b-$\nhi~distribution in both simulations
nicely converges to the distributions in the IGM at large impact parameters 
($r_{\perp} > 10 $ pMpc). 
Bottom panels of Fig.~\ref{fig.kde_large_impact} compare $b-$\nhi~distributions
from forward modelled \lya~forest spectra. Similar to lower and intermediate 
impact parameters (bottom panels of Fig.~\ref{fig.kde_small_impact} and
Fig.~\ref{fig.kde_intermediate_impact}) the $b-$\nhi~distributions show 
that most of the \lya~absorption lines are at higher column densities 
($\log (N_{\rm HI}/ \text{cm}^{-2}) >13$) spanning a limited range in $b$ in both simulations. 
The distribution shapes and even the normalization indicated by the color bar look 
similar. This shows that the differences seen in the $b-$\nhi~distributions 
obtained with both perfect and forward modeled \lya~spectra are washed out 
at these large impact 
parameters ($\sim 10 $ pMpc) from massive halos. Also the $b-$\nhi~distribution 
is in very good agreement with the one seen in the case of IGM 
(see \citetalias{Khaire23}) suggesting that at high-impact parameters we 
are probing the gas residing mainly in the IGM (also see Fig.~\ref{fig.bn_kde_apendix0} 
and \ref{fig.bn_kde_apendix1} in Appendix).

As shown in the \citetalias{Khaire23}, the shape of the 
$b-$\nhi~distributions for IGM obtained 
from perfect \lya~forest spectra (S/N $=100$ per pixel and $\Delta v = 4.2$ km s$^{-1}$) 
can distinguish the effect of feedback. The same is true for the shape of the 
$b-$\nhi~distributions around massive halos for impact parameters of 
$\gtrsim 1.5$ pMpc. This signal in shape, however, gets washed out in forward-modeled
\lya~spectra but the number of 
\lya~absorption lines show significant differences till impact parameters 
of $\sim 10$ pMpc in both perfect as well as forward modeled \lya~forest. 

Given that the main difference in  $b-$\nhi~distributions between two simulations 
from forward modeled \lya~forest is in the number of  \lya~absorption lines, in 
Fig.~\ref{fig.halo_forward_hist_0} we plot the individual histograms of $b$ (left-hand 
panels) and $N{\rm HI}$ (right-hand panels) at the same impact parameter bins for which 
the $b-$\nhi~distributions are plotted in Fig.~\ref{fig.kde_small_impact},
\ref{fig.kde_intermediate_impact} and \ref{fig.kde_large_impact}.
These histograms are plotted from  $\sim 10^5$ forward modeled \lya~forest spectra
around halos in each simulation. At all impact parameter bins, the shapes of both 
$b$ and \nhi~histograms are identical but there is a clear difference in the amplitude
showing the difference in the number of \lya~absorption lines detected by VPfit. 
For reference, the number of \lya~lines plotted in the histogram is indicated on the 
right-hand panel of Fig.~\ref{fig.halo_forward_hist_0}. At small and intermediate impact 
parameter bins (top and middle panel), the number of \lya~lines in IllustrisTNG is higher 
than Illustris by a factor of $\sim 2$. Whereas at large impact parameters of $10$ pMpc 
(bottom panel) IllustrisTNG shows only 18 percent more \lya~lines than Illustris. 
To further investigate the trend in the number of \lya~lines
detected, we show individual histograms of $b$ and \nhi~distributions at different 
impact parameters in  Fig~\ref{fig.halo_forward_hist_1} and \ref{fig.halo_forward_hist_2} 
in the Appendix. These plots and the difference in the number of lines detected in each 
bin illustrate a trend where there are large differences in the
detected \lya~lines at lower impact parameters and with increasing impact parameter
this difference diminishes. This is expected since at high-impact parameters from halos we
are probing the regions in the IGM and we expect the number of lines for the same 
redshift path length should be the same since we calibrated both simulations with
$\Gamma_{\rm HI}$ values via matching the ${\rm dN/dz}$. 

Given the prominent difference in the number of \lya~lines detected as a function 
of impact parameter, we now focus on the simple statistics of line density profile around 
massive halos ${\rm dN/dz}\,(r_{\perp})$, as discussed in the following subsection.

\subsection{The line density profile around halos}\label{sec.dndz}
The line density profile around halos, ${\rm dN/dz}\,(r_{\perp})$, is the 
number of \lya~absorption lines within some \nhi~range 
per unit redshift pathlength as a function of $r_{\perp}$. 
In our forward models, we are sensitive to the \nhi~$\sim 10^{12.5}$ cm$^{-2}$,  and the 
lines above \nhi~$> 10^{14.5}$ cm$^{-2}$ are usually saturated. However,
to be consistent with \nhi~range that is used for the analysis in \citetalias{Khaire23}
and for tuning the values of $\Gamma_{\rm HI}$ here (in section \ref{sec.gamma}),
we choose the range of  $10^{12}<$ \nhi~$<10^{14.5}$ cm$^{-2}$ 
for ${\rm dN/dz}\,(r_{\perp})$ calculation.
At each impact parameter bin centered at $r_{\perp}$ we count all 
the \lya~lines ($N_{Ly\alpha}$) within the velocity range $\Delta {v_{z}} = \pm 500$ 
km  s$^{-1}$ as identified by VPfit falling in our column density range. 
We calculate the total path length $\Delta z$ covered by the 
sightlines in each impact parameter bin by multiplying the number of 
sightlines in the bin ($N_{\rm fit}$) by the redshift path of a 
single sightline i.e. the redshift path $\Delta z_{1000}$ 
for the central 1000 km s$^{-1}$ segment of the sightlines since
we only count lines in the velocity range $\Delta {v_{z}} = \pm 500$ km 
s$^{-1}$. For this calculation, we use the same 
cosmological parameters with which the simulations are run
and measure the line density at each impact parameter bin as
${\rm dN/dz} = N_{Ly\alpha} / (N_{\rm fit} \, \Delta z_{1000})$. 
In this way, we calculate ${\rm dN/dz}$ at different $r_{\perp}$
bins from 0.1 to 90 pMpc\footnote{We chose 23  
$r_{\perp}$ bins of variable sizes to accommodate a sufficiently large number of sightlines
(up to $10^{5}$) to have precise ${\rm dN/dz}$.}
in Illustris and IllustrisTNG.

In order to compute the ${\rm dN/dz}$ from ambient IGM we 
generate forward-modeled IGM sightlines. We follow the same
forward modeling procedure as in the case of halo sightlines as explained in
Section~\ref{sec.fwd_halos}. These are randomly drawn sightlines and
we do not restrict to a velocity window, rather we use full sightlines. As expected we find 
${\rm dN/dz}$ values for the IGM (with the same $10^{12}<$ \nhi~$<10^{14.5}$ cm$^{-2}$ range)
for both Illustris (${\rm dN/dz}=71.5$) and IllustrisTNG (${\rm dN/dz} = 72.3$) 
to be essentially the same, i.e. the difference is within 1.1\%. 
This is expected because we have calibrated $\Gamma_{\rm HI}$ by matching 
${\rm dN/dz}$ of IGM (see Section~\ref{sec.gamma}). 
However, note that the ${\rm dN/dz}$ values quoted here ($71.5$ and 
$72.3$) are different from the one we used (${\rm dN/dz} = 205$) from \citetalias{Khaire23}
to calibrate the $\Gamma_{\rm HI}$ using \citet{Danforth14} data (Section~\ref{sec.gamma})
because the forward models are different in both cases. Here we use S/N $=7$
per pixel (of 6 km s$^{-1}$; 12.2 per resolution element) same for all sightlines 
but for \citet{Danforth14} data we had different S/N 
in each quasar spectrum. For \citet{Danforth14} data we used the condition 
S/N$>17$ per resolution element (i.e S/N $>10$ per pixel of 6 km s$^{-1}$), 
same as done for \lya~forest power spectrum calculation in \citet{Khaire19}. 
It is reassuring that our calibrated $\Gamma_{\rm HI}$ results
in same ${\rm dN/dz}$ for both simulation irrespective of the S/N of data 
used for calibration.

\begin{figure}
\includegraphics[width=0.48\textwidth,keepaspectratio]{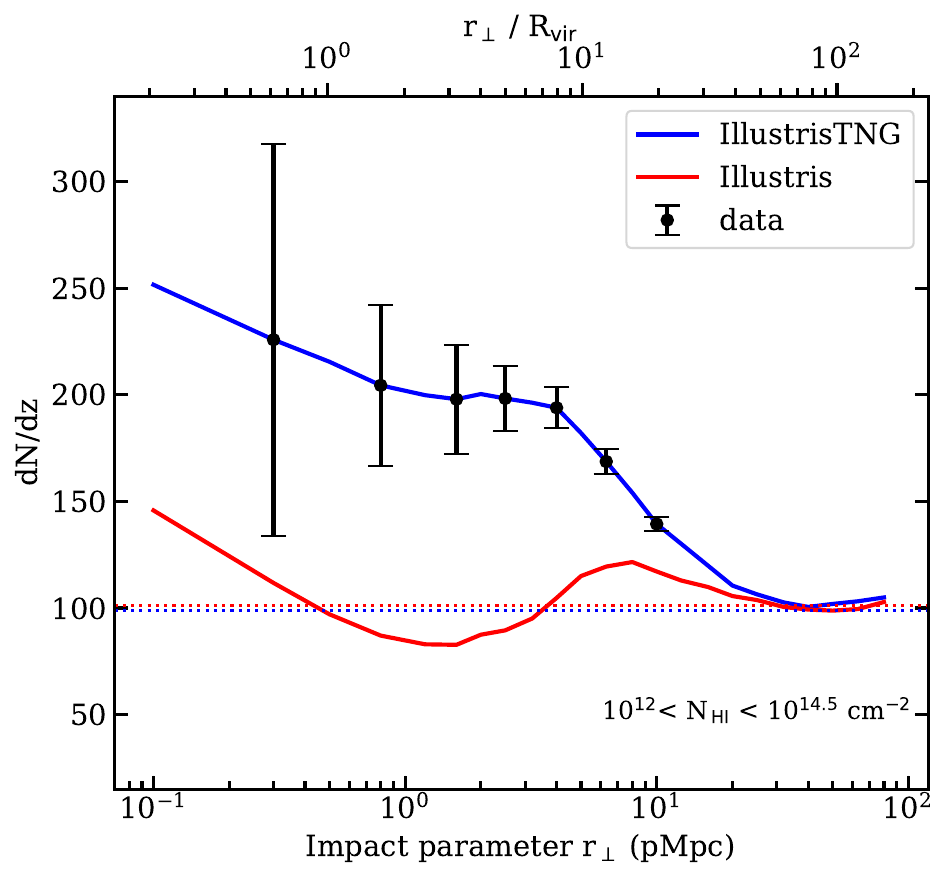}
\caption{The line density profile 
${\rm dN/dz} (r_{\perp})$ for $10^{12} <N_{\rm HI} < 10 ^{14.5}\, { \rm cm^{-2} }$ absorbers 
obtained from 
forward modeled mock spectra around LRG host 
halos within $\pm 500~{\rm km~s^{-1}}$ along line-of-sights. 
IllustrisTNG (blue) gives a factor of $\sim 1.5-2$ higher 
${\rm dN\slash dz}$ as compared to Illustris (red) 
out to $r_\perp\lesssim$ $5$ $\,{\rm pMpc}$ ($\sim 10 \, R_{\rm vir}$ ). 
By construction, both simulations converge to the expected IGM ${\rm dN/dz}$ 
(dotted lines) at large distances from halos.  
The black points with error bars show the expected precision using
archival data assuming IllustrisTNG is the true model (see 
section~\ref{sec.data}
for more details). With this data, we can distinguish between 
Illustris and IllustrisTNG 
feedback models at 12$\sigma$ statistical significance.}
\label{fig.dndz}
\end{figure}

\begin{figure*}
\includegraphics[width=0.98\textwidth,keepaspectratio]{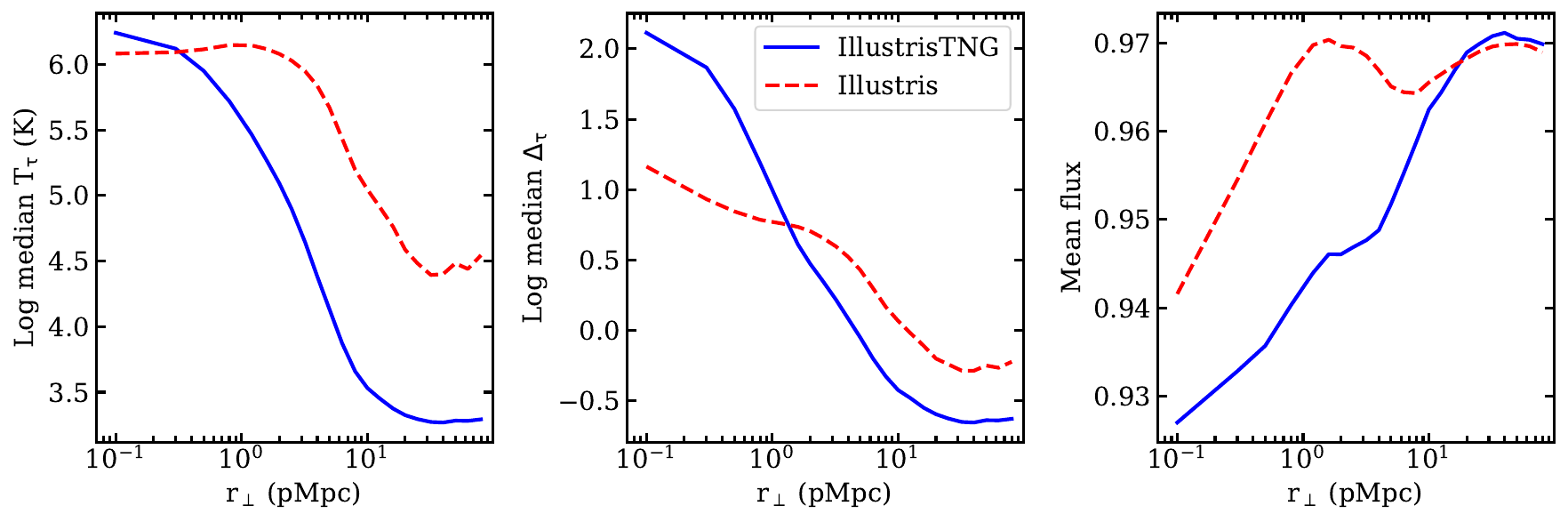}
\caption{Median of optical depth weighted temperature ${\rm T}_{\tau}$ 
(left-hand panel) and overdensity 
$\Delta_{\tau}$ (middle panel), and mean \lya~flux (right-hand panel) for the line of sights 
passing at different impact parameters ($r_{\perp}$)
along $\pm 500$ km $s^{-1}$ of the position of halos for Illustris (dashed curves) and 
IllustrisTNG (solid curves). The differences in all three
quantities help explaining the trend in ${\rm dN/dz} \, (r_{\perp})$ 
seen in Fig.~\ref{fig.dndz} (as discussed in the Section~\ref{sec.dndz}).}
\label{fig.temp_rho_flux}
\end{figure*}

In Fig.~\ref{fig.dndz} we plot the  ${\rm dN/dz} (r_{\perp})$ from the forward 
modelled \lya~forest in both simulations up to $r_{\perp} =  90 $ pMpc.
The top axis shows $r_{\perp}/R_{\rm vir}$ where we use a representative value of
$R_{\rm vir} = 0.49$ pMpc obtained for the $z=0.1$ halo with  
mass $10^{12.9}$ M$_{\odot}$ which is the threshold $M_{\rm min}$
for IllustrisTNG LRG hosts. 
To understand the interesting trends seen in ${\rm dN/dz} (r_{\perp})$,
in Fig.~\ref{fig.temp_rho_flux} we plot the median values of the optical depth weighted 
temperature ($T_{\tau}$) and over-densities ($\Delta_{\tau}$), and mean flux  
at different $r_{\perp}$ calculated over the same $\pm 500$ km s$^{-1}$ window. 
As expected, both simulations show that ${\rm dN/dz}$ (in Fig.~\ref{fig.dndz}) 
to be higher near halos and it decreases monotonically at larger $r_{\perp}$ up to 
$1 $ pMpc (i.e $\sim 2 R_{\rm vir}$).
This has been ubiquitously seen in the CGM observations in various surveys 
\citep[e.g,][]{Werk12, Chen18}.
The decline in $\rm dN/dz$ seen here is mainly because of the decline in the gas density
with $r_{\perp}$ as shown in the middle panel of Fig.~\ref{fig.temp_rho_flux}.
Whereas, the temperatures are almost the same up until 
$\sim 2 R_{\rm vir}$ in both simulations (left panel of Fig.~\ref{fig.temp_rho_flux}) 
because of the viral shocks near halos. 
This is more evident when we compare the decline in median $T_{\tau}$ and $\Delta_{\tau}$
in both simulations. For example, 
Fig.~\ref{fig.temp_rho_flux} shows that 
up to $r_{\perp} =1 $ pMpc,  the median $\Delta_{\tau}$ drops by an order of 
magnitude in the case of IllustrisTNG whereas the drop in median $T_{\tau}$
is of the factor of only $\sim 3-4$. In the case of the Illustris simulation,
median $\Delta_{\tau}$ drops by a factor of $\sim 3$ (i.e 0.5 dex) but 
the median $T_{\tau}$ is essentially the same up until $r_{\perp} \simeq 2$ pMpc. 
Therefore the density profile of halos mainly causes the decline of 
${\rm dN/dz}$ at low $r_{\perp}$ seen up to $\sim 1$ pMpc. 

At higher $r_{\perp}$ ($>3$ pMpc for Illustris and  $>1$ pMpc for IllustrisTNG)
the median $T_{\tau}$ starts to drops with increasing $r_{\perp}$ which 
compensates the decreasing $\Delta_{\tau}$ (see Fig.~\ref{fig.temp_rho_flux}). 
Depending of the rate of change of $T_{\tau}$ and $\Delta_{\tau}$ with $r_{\perp}$ 
either the ${\rm dN/dz} (r_{\perp})$ flattens, 
as in the case of IllustrisTNG ($1 < r_{\perp}<5$ pMpc; Fig.~\ref{fig.dndz})
or ${\rm dN/dz} (r_{\perp})$ increases 
as in the case of IllustrisTNG ($2 < r_{\perp}<10$ pMpc; Fig.~\ref{fig.dndz}). 

At very high impact parameters ($r_{\perp} > 20$ pMpc) median 
$T_{\tau}$ and $\Delta_{\tau}$ 
remains constant with $r_{\perp}$ (Fig.~\ref{fig.temp_rho_flux}) 
as they converge to  $T_{\tau}$ and $\Delta_{\tau}$ for the IGM. 
This results in a constant ${\rm dN/dz} (r_{\perp})$ for both simulations. 
Also the values of this asymptotic ${\rm dN/dz}$ are the same in both simulations
and for IGM. The ${\rm dN/dz}$ for IGM is 
shown by dotted lines in Fig.~\ref{fig.dndz} to which ${\rm dN/dz} (r_{\perp})$ 
from both simulations converge at high $r_{\perp} > 20$ pMpc. 

Interestingly, in the case of
Illustris the ${\rm dN/dz}$ goes below the ${\rm dN/dz}$ of IGM for 
a range of intermediate impact parameters $0.7 < r_{\perp} < 3$ pMpc. This is mainly
because Illustris expels a huge amount of gas from the CGM of galaxies into the IGM. 
This is why the median $\Delta_{\tau}$ (see Fig~\ref{fig.temp_rho_flux}) for Illustris
is lower than the IllustrisTNG (by an order of magnitude) near halos ($r_{\perp}  < 1$ pMpc)
and higher than IllustrisTNG at large impact parameters  ($r_{\perp} > 1$ pMpc). 
The constant offset between $\Delta_{\tau}$ at higher impact parameters  ($r_{\perp} > 10$ pMpc)
arises because of higher   $T_{\tau}$  in Illustris requiring the higher overdenisties to
produce the \lya~absorption. 
The mean flux shown in (right-hand panel of) Fig.~\ref{fig.temp_rho_flux} shows that
the mean flux for Illustris rises beyond the mean flux of the IGM at intermediate $r_{\perp}$ 
($0.8 < r_{\perp} < 20 $ pMpc) providing lower than IGM ${\rm dN/dz}$ (see Fig.~\ref{fig.dndz}) 
and then it drops to give a rise in  ${\rm dN/dz}$ as seen in Fig.~\ref{fig.dndz}. 
However, even though
asymptotic values of $T_{\tau}$ and $\Delta_{\tau}$ at largest impact parameters
in both simulations are different, mean \lya~flux in  both 
simulations converge to the mean flux of the IGM (right-hand panel of 
Fig.~\ref{fig.temp_rho_flux}) because of our calibrated $\Gamma_{\rm HI}$. 

As shown in Fig.~\ref{fig.dndz}, Illustris and IllustrisTNG simulations 
provide very different ${\rm dN/dz}(r_{\perp})$ profile around massive halos.
Up to an impact parameter of 5 pMpc i.e $\sim 10 \,  R_{\rm vir}$, the
${\rm dN/dz}(r_{\perp})$ obtained for IllustrisTNG is a factor of 
$1.5-2$ times higher than the one obtained for Illustris. 
This is the consequence of the different feedback prescriptions employed 
in the simulations. 
The difference in ${\rm dN/dz}(r_{\perp})$ can be seen until $r_{\perp} \sim 20$ pMpc
($\sim 40 R_{\rm vir}$). 
This striking difference in the ${\rm dN/dz} (r_{\perp})$ suggests that it 
will be a powerful tool to study the impact of AGN feedback. 
Moreover, the observed ${\rm dN/dz} (r_{\perp})$ determined from HST-COS data 
at low-$z$ can be a benchmark observation against which simulations can test
and calibrate their feedback models. Motivated by this, 
in what follows we perform a feasibility analysis to investigate if the currently 
available HST-COS quasar spectra probing massive halos hosting LRGs 
have constraining power to distinguish feedback models in Illustris and IllustrisTNG. 

\subsection{The Feasibility of using the Line Density Profile as a Probe of the Feedback} \label{sec.data}

We have performed forward models taking into account the S/N and redshift distribution of
set of 94 HST-COS background quasar spectra that
probe 3193 foreground LRGs (see Section~\ref{sec.data} and Fig.~\ref{fig.data_archival}).
Therefore for the feasibility analysis, we use $r_{\perp}$ distribution of LRGs 
(right-hand panel of Fig.~\ref{fig.data_archival}) to estimate the precision one can obtain on 
the ${\rm dN/dz} (r_{\perp})$.
For that, we re-bin the impact parameter distribution of these LRGs in logarithmic bins.
We chose the bin centers such that they match the centers of our impact parameter
bins where we have performed previous analyses in the simulations.
We determine the number $N^{\rm bin}$ of LRGs in each impact parameter bin. 
With this information, for every impact parameter bin
we randomly choose $N^{\rm bin}$ forward-modelled halo sightlines 
falling in the corresponding impact parameter bin of the simulations. 
By doing this at each impact parameter bin, we are choosing a forward model 
\lya~sightlines for each LRG-quasar pair 
passing within the respective $(r_{\perp})$ bin in simulations. 
We call this as our first random sample. We generate such $2000$ random samples.
We generate these samples only using IllustrisTNG simulation thereby assuming 
it to be the underlying true model. 
Since we have already run VPfit on forward modeled \lya~spectra, for each sample
we determine the \nhi~values within the same velocity window of $\pm 500$ km s$^{-1}$ 
$N_{\rm HI}$ range and 
estimate the ${\rm dN/dz}$ for absorption lines 
and calculate the sample average of ${\rm dN/dz}$ and its variance. 

We show these
mock measurements i.e the sample average ${\rm dN/dz}$ and its variance, 
in Fig.\ref{fig.dndz} by black data points with error bars. These mock measurements 
are the result of an ensemble of $2000$ measurements. 
As expected the sample mean of ${\rm dN/dz}$ falls on the curve representing
the true model i.e. the IllustrisTNG simulation.
Since there are a large number of absorbers per bin we can find the 
statistical significance $s$ of these mock measurement to distinguish ${\rm dN/dz}$ 
of IllustrisTNG (the true model) from Illustris using the following equation,
%
\begin{equation}
    s^2 = \sum_{i} \big[\Delta ({\rm dN/dz})_i/\sigma_i\big]^2,
\end{equation}
where $i$ denotes $i^{\rm th}$ bin of impact parameter at which the 
mock measurements are done and 
the $\Delta ({\rm dN/dz})_i$ is the difference between the 
${\rm dN/dz}$ of Illustris and our mock measurement in the same bin.
We find that the mock  ${\rm dN/dz}$ measurements can distinguish IllustrisTNG from Illustris
simulation with a statistical significance of 12$\sigma$. 
Note that this is not actually 
the statistical significance of any given mock dataset but of an ensemble of datasets, since 
for a single dataset, there would be fluctuations. 

This simple analysis reveals the potential of using the line density profile
around halos ${\rm dN/dz} \, (r_{\perp})$ to constrain different feedback models used in 
galaxy formation simulations. Most of the feedback-constraining observations are related
to the properties and distribution of galaxies whereas ${\rm dN/dz} \, (r_{\perp})$
uses the distribution of gas around galaxies up to 10 pMpc. 
Therefore, ${\rm dN/dz} \, (r_{\perp})$ holds considerable promise as it offers 
a complementary constraint for assessing the AGN feedback.

\section{Summary}\label{sec.summary}
Galaxy formation simulations rely on AGN feedback to regulate star formation in massive galaxies, 
a process involving powerful outflows and the heating of the surrounding gas 
in order to regulate star formation. However, these 
simulations cannot resolve the central AGNs directly, necessitating the use of sub-grid 
implementations for feedback modeling. While various feedback prescriptions are employed in 
simulations, their parameters are typically tuned based on observations, most of which pertain to 
galaxy properties and the immediate vicinity of galaxies within approximately the virial radius.
What is lacking are benchmark observations that extend beyond galaxies, reaching into the IGM
to probe and constrain these feedback models. This is a notable gap to address, given the relative 
ease of reproducing the IGM in simulations based on well-established theoretical foundations. 
Therefore, in this work, we seek summary statistics based on observations 
that are capable of probing the gas surrounding massive galaxies at vast distances with 
the potential to serve as a critical constraint for AGN feedback modeling. 

To achieve this objective, we utilized two state-of-the-art galaxy formation 
simulations; Illustris and IllustrisTNG at $z=0.1$. These simulations are 
run with the same initial conditions and 
almost identical codes but with different feedback implementations. 
A notable difference is in the implementation of the radio mode AGN feedback which is very 
aggressive in the Illustris as compared to IllustrisTNG. As a result,
the massive halos in the Illustris simulation generate large
shock-heated regions surrounding them (see Fig.~\ref{fig.boxplot}) extending up to 10 pMpc.
Moreover, Illustris expels large amounts of gas from the halos of galaxies. 
It was in fact one of the key observations that prompted different AGN feedback 
prescriptions in the IllustrisTNG 
\citep[see][]{IllustrisTNG, Pillepich18}. As a result of different AGN feedback prescriptions,
these two simulations are ideal for testing the effect of AGN feedback on the properties 
of gas around halos and in the IGM. 

Because the fraction gas in the diffuse low-temperature gas that is responsible for
the observed \lya~absorption is quite different in 
both simulations ($23.2$\% in Illustris and $38.5$\% in IllustrisTNG), 
we perform a calibration of the UV background in each simulation. 
This calibration involves adjusting $\Gamma_{\rm HI}$ to ensure that both simulations yield
the same observed \lya~line density ${\rm dN/dz}$ for the IGM. 
This calibration process, as outlined in \citetalias{Khaire23}, is essential to facilitate 
meaningful comparisons between the simulations and observational data.

After tuning the $\Gamma_{\rm HI}$ in both simulations we select the halos
that are potential hosts of LRGs and generate perfect (S/N $=100$ with 
resolution $4.1$ km s$^{-1}$) and mock HST COS \lya~forest 
spectra (S/N $=7$ per pixel of $6$ km s$^{-1}$; 12 per COS resolution element) 
at different impact parameters $r_{\perp}$. 
We fit these spectra using our automatic Voigt profile code and study statistics such
as 2D $b-$\nhi~distributions, $b$~distributions, \nhi~distributions and ${\rm dN/dz}$
with ($r_{\perp}$). 

For perfect \lya~forest spectra very close to the halos within  $\sim 2 r_{\rm vir}$, 
the shape of 2D $b-$\nhi~distributions is similar in both simulations and differs at 
higher impact parameters. (see Fig.~\ref{fig.kde_small_impact}, 
\ref{fig.kde_intermediate_impact}, ~\ref{fig.kde_large_impact} \& \ref{fig.bn_kde_apendix0}).
Lines move towards low $b$ values and higher \nhi~with increasing $r_{\perp}$. At very high
impact parameters ($r_{\perp} > 10 $ pMpc) $b-$\nhi~distributions slowly converge to
the $b-$\nhi~distributions of the IGM. 
But in forward modeled \lya~forest, the differences in the shape 
of $b-$\nhi~distributions diminish at all impact parameters. 
However, there is a clear difference in the
normalization of 2D $b-$\nhi~distributions, which is proportional to the number of 
\lya~lines (see Fig.~\ref{fig.halo_forward_hist_0},~\ref{fig.halo_forward_hist_1} 
\&~\ref{fig.halo_forward_hist_2}). Motivated by this we calculate the 
${\rm dN/dz} \, (r_{\perp})$, i.e. \lya~line density profile, around halos for 
both Illustris and IllustrisTNG using forward modeled \lya~spectra. 
We find that ${\rm dN/dz}\, (r_{\perp})$ obtained for
IllustrisTNG is larger than Illustris by a factor of $1.5-2$ up to an 
impact parameter of 10 pMpc ($\sim 20$ viral radii; Fig~\ref{fig.dndz}).
This significant difference in ${\rm dN/dz} \, (r_{\perp})$ underscores its 
promising utility for probing and constraining diverse feedback implementations 
used in simulations

Lastly, we perform feasibility analysis using available archival data. 
We find that there are 3193 foreground LRGs probed by high-quality 
94 HST COS background quasar spectra 
up to an impact parameter of 10 pMpc  (see Fig.~\ref{fig.data_archival}). 
Using their real impact parameter distribution and median S/N of the 
sample, we estimate precision on the mock ${\rm dN/dz} \, (r_{\perp})$ measurement 
(see Fig.~\ref{fig.dndz}). We find that these mock measurements can differentiate
between Illustris and IllustrisTNG simulations 
with a statistical significance of 12$\sigma$.
This demonstrates that the ${\rm dN/dz} (r_{\perp})$ measurement around massive halos, derived 
from the presently available HST-COS data, holds the capacity to discern between extreme feedback 
models. Moreover, ${\rm dN/dz} (r_{\perp})$ can serve as a valuable benchmark observation for 
rigorously testing and calibrating various feedback prescriptions employed in simulations.

\section*{acknowledgement} 
VK thanks D. Sorini and D. Nelson for helping with Illustris simulation datasets.
VK thanks R. Srianand for hosting him at the Inter-University Centre for Astronomy 
and Astrophysics (IUCAA), Pune, India, and for helpful discussions on the paper. 
VK is supported through the INSPIRE Faculty Award (No.DST/INSPIRE/04/2019/001580) 
of the Department of Science and Technology (DST), India.

\section*{Data Availability}
We utilized a subset of quasar spectra from the dataset presented in 
\citet{Danforth14} which is accessible at 
\href{https://archive.stsci.edu/prepds/igm}
{https://archive.stsci.edu/prepds/igm}. The simulation data is obtained from the 
Illustris website, with links provided in the text at relevant locations.

\bibliographystyle{mnras}
\bibliography{vikrambib}

\appendix
\section{The 2D and marginalized distribution of Doppler parameter and \nhi~around halos}
\begin{figure*}
\begin{subfigure}{.49\textwidth}
  \centering
  \includegraphics[width=\linewidth,keepaspectratio]{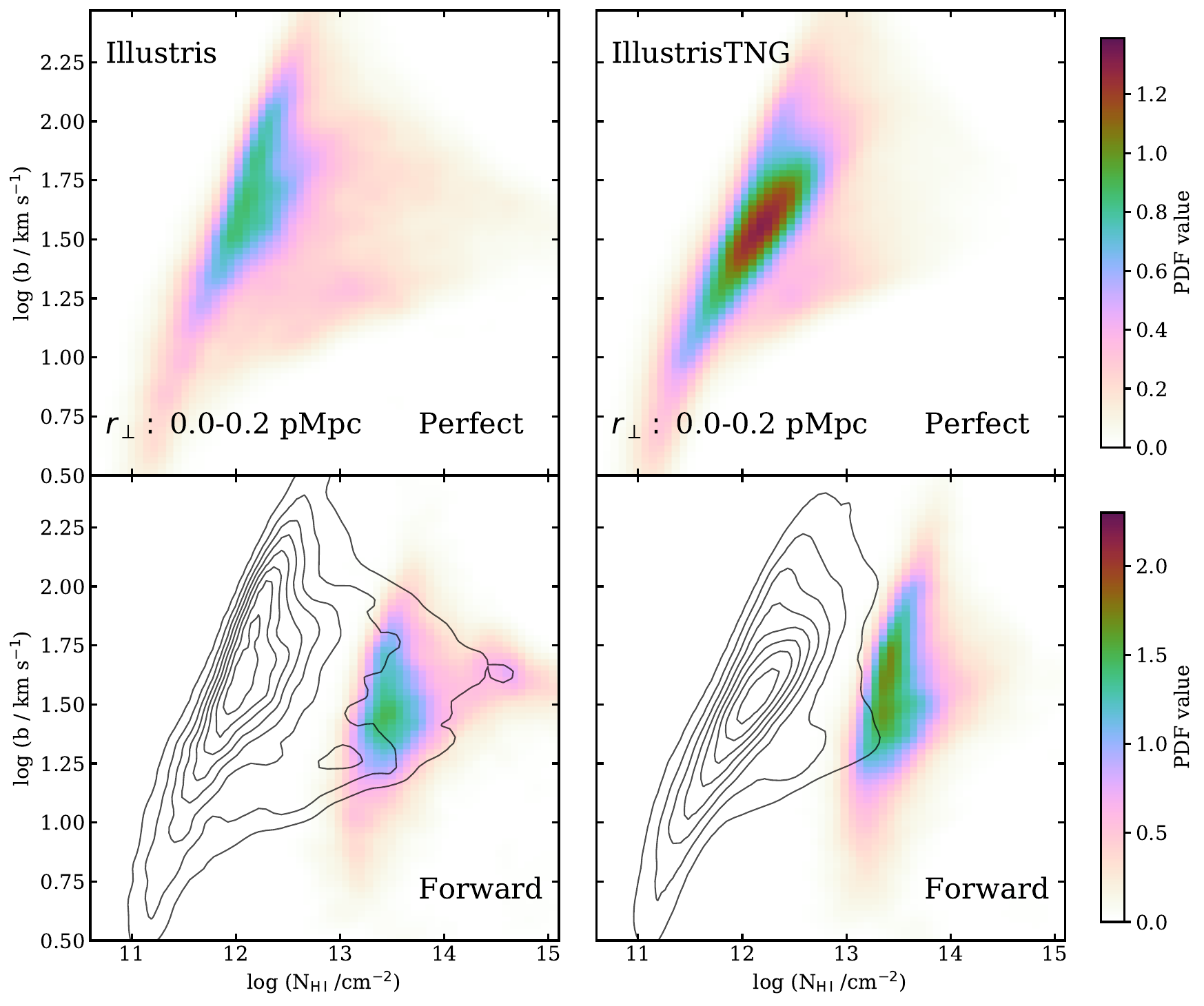}  
\end{subfigure}
\begin{subfigure}{.49\textwidth}
  \centering
  \includegraphics[width=\linewidth,keepaspectratio]{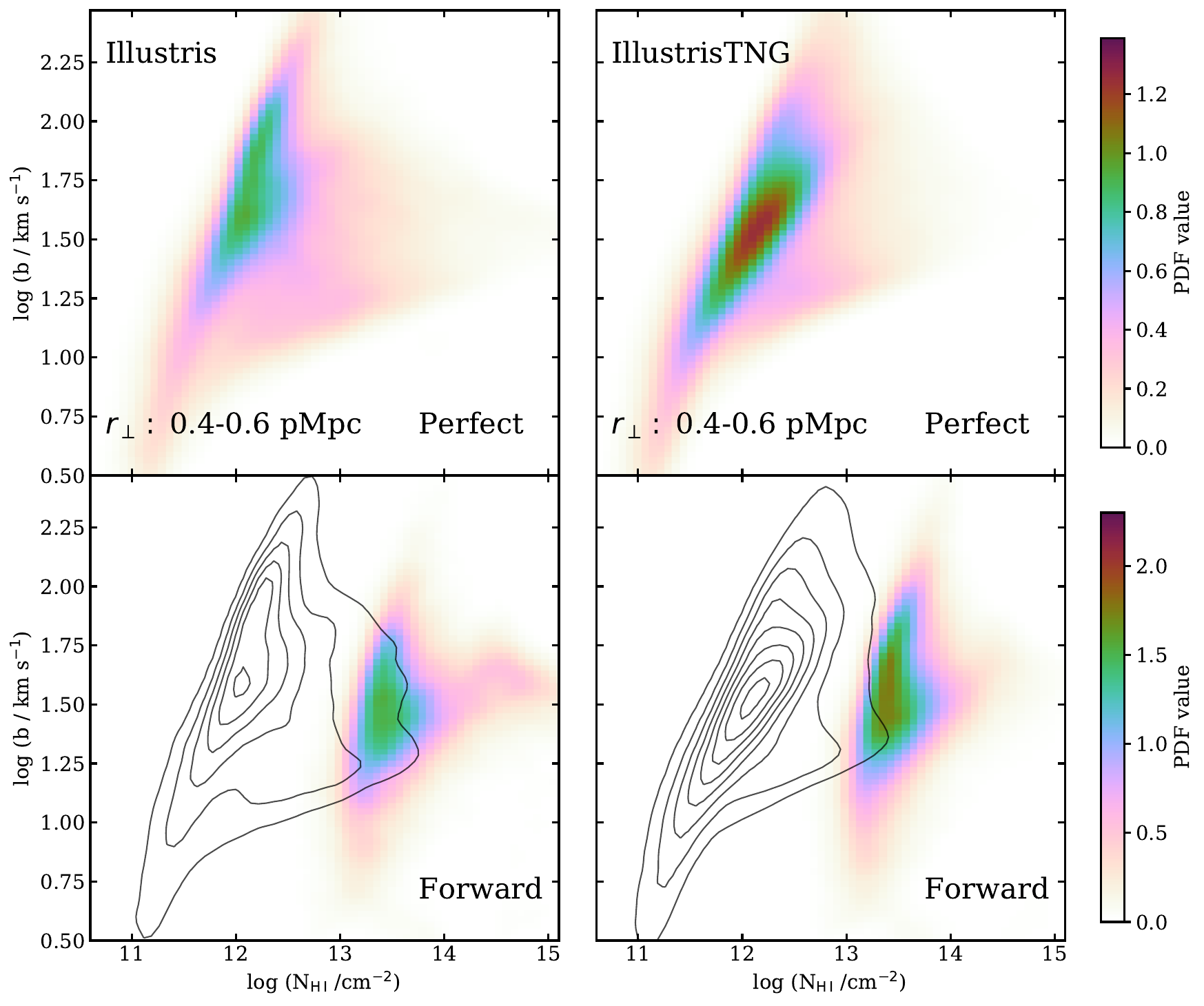}  
\end{subfigure}

\begin{subfigure}{.49\textwidth}
  \centering
  \includegraphics[width=\linewidth,keepaspectratio]{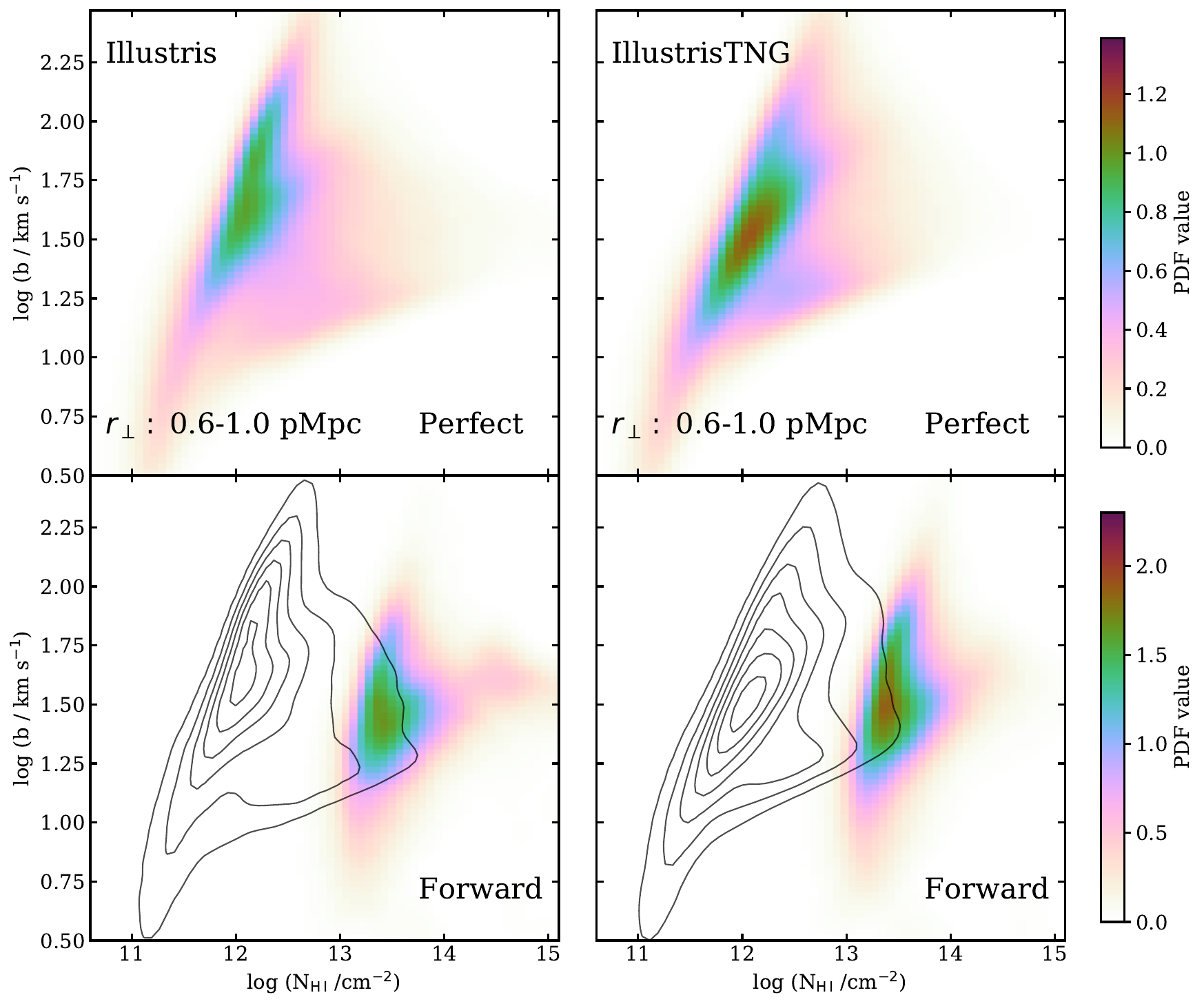}  
\end{subfigure}
\begin{subfigure}{.49\textwidth}
  \centering
  \includegraphics[width=\linewidth,keepaspectratio]{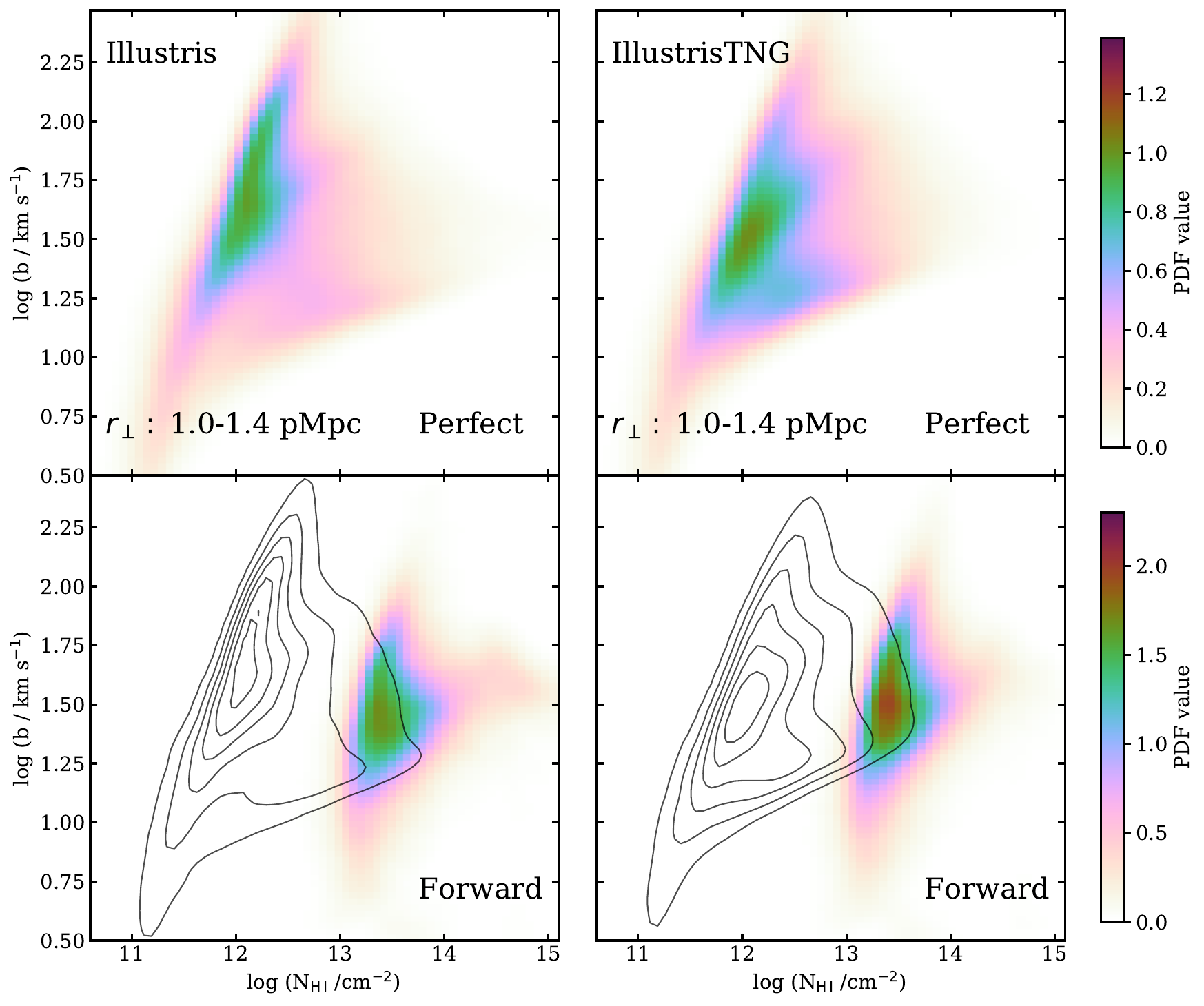}  
\end{subfigure}

\begin{subfigure}{.49\textwidth}
  \centering
  \includegraphics[width=\linewidth,keepaspectratio]{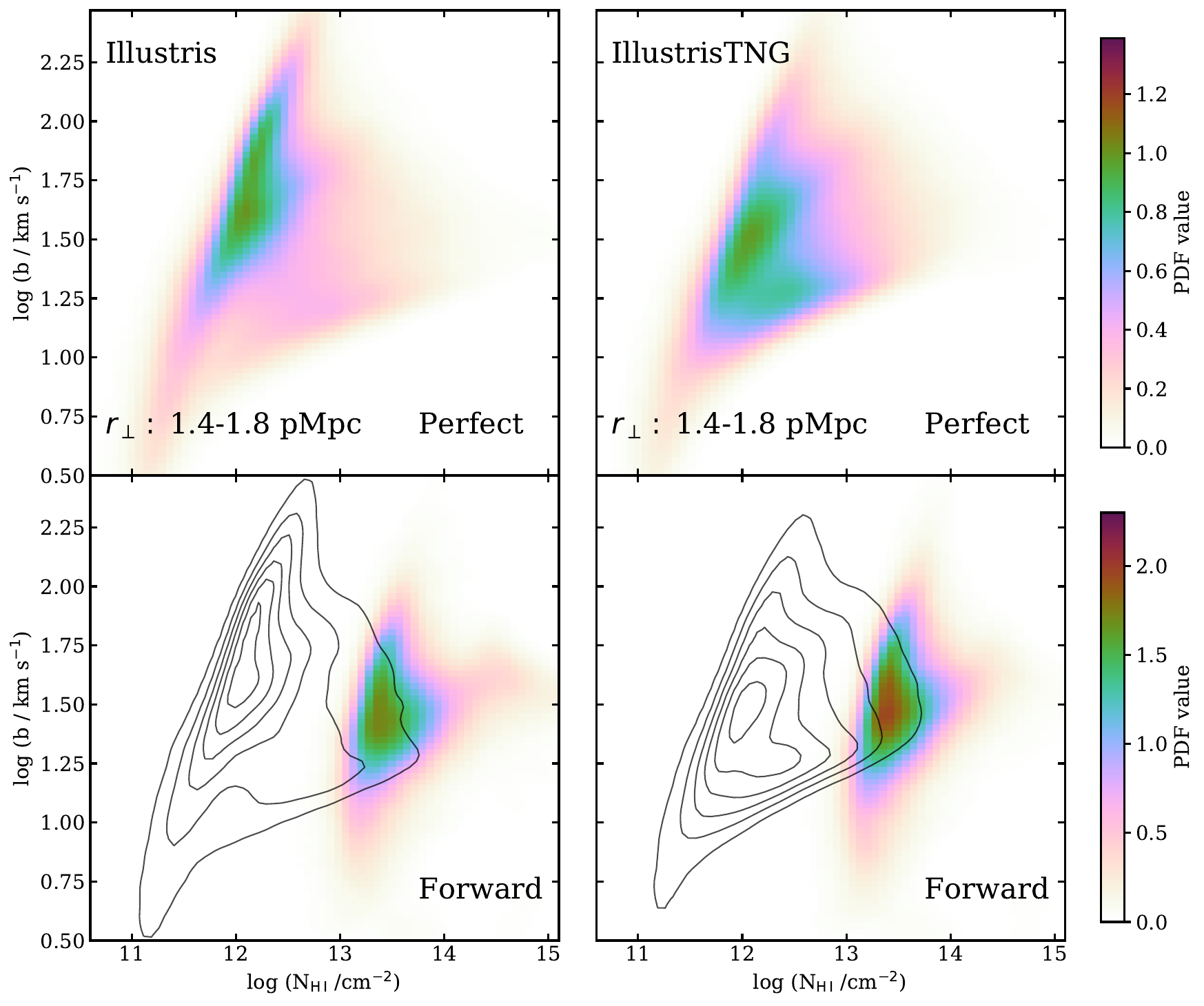}  
\end{subfigure}
\begin{subfigure}{.49\textwidth}
  \centering
  \includegraphics[width=\linewidth,keepaspectratio]{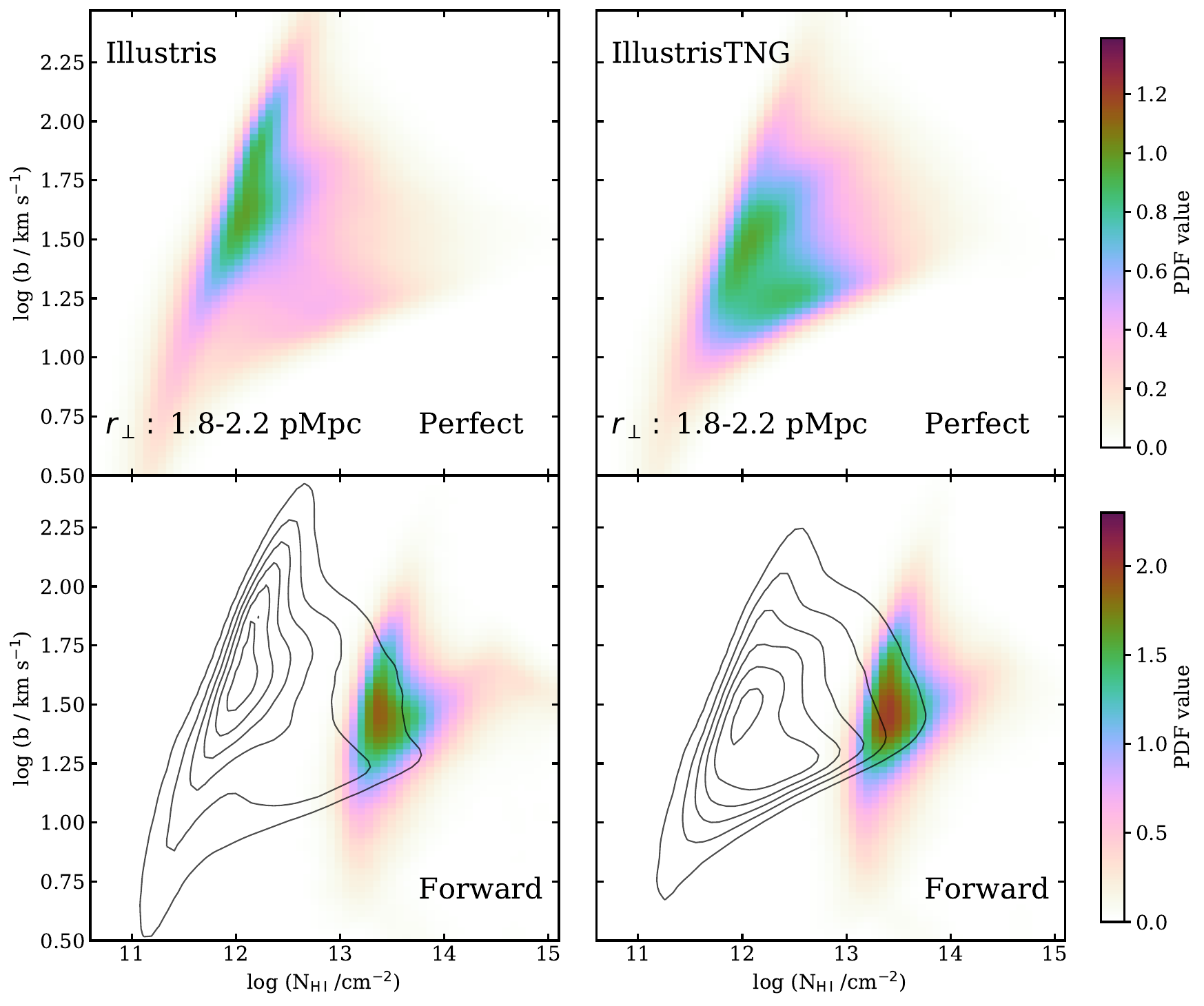}  
\end{subfigure}

\caption{Same as Fig.~\ref{fig.kde_small_impact}, the 2D $b$-\nhi~distribution around massive halos at six 
different impact parameters (indicated in legends) from 0 to 2.2 pMpc. With increasing impact parameters the shape and normalization (indicated by color) of the
distribution evolves. }
\label{fig.bn_kde_apendix0}
\end{figure*}

\begin{figure*}
\begin{subfigure}{.49\textwidth}
  \centering
  \includegraphics[width=\linewidth,keepaspectratio]{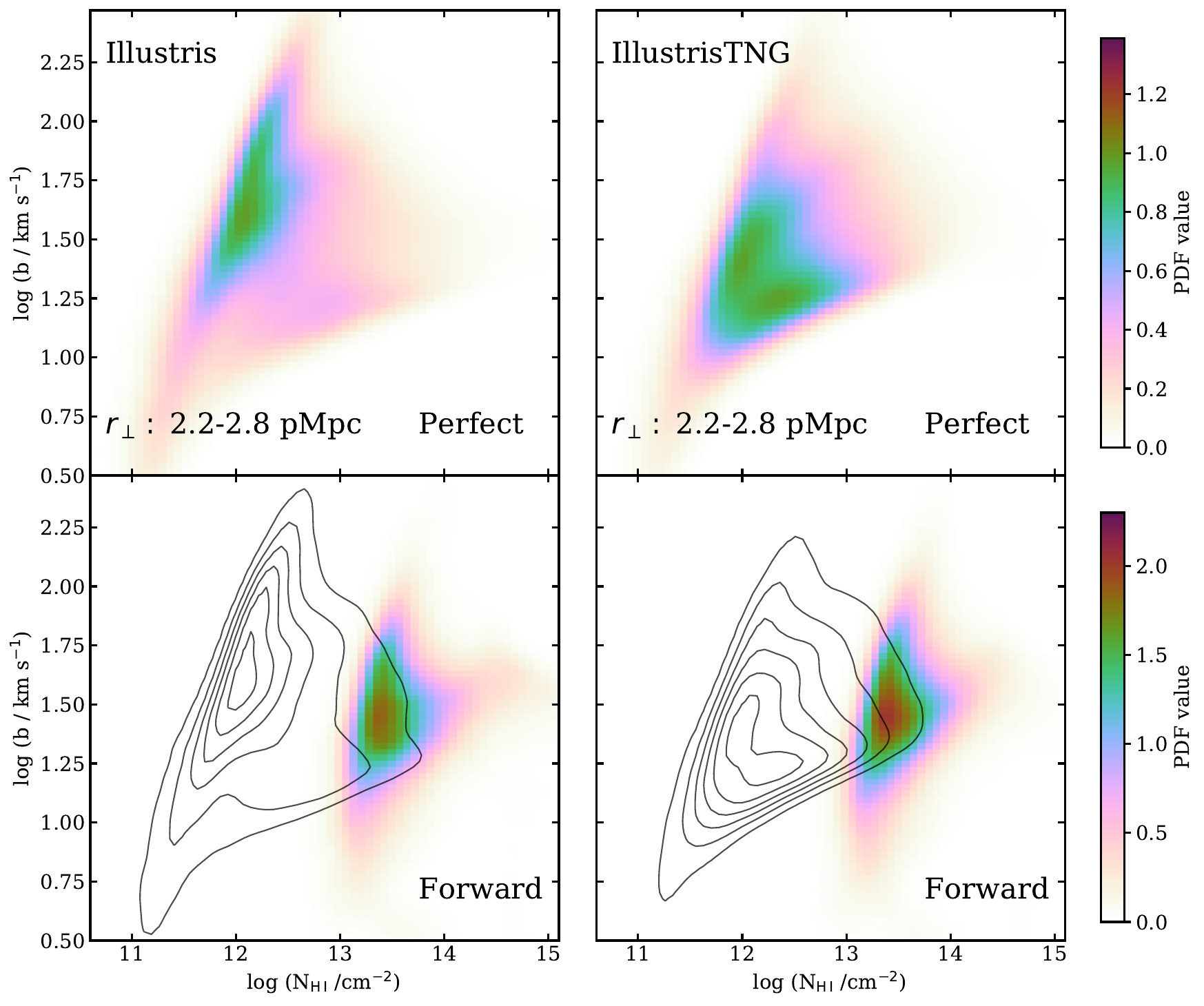}  
\end{subfigure}
\begin{subfigure}{.49\textwidth}
  \centering
  \includegraphics[width=\linewidth,keepaspectratio]{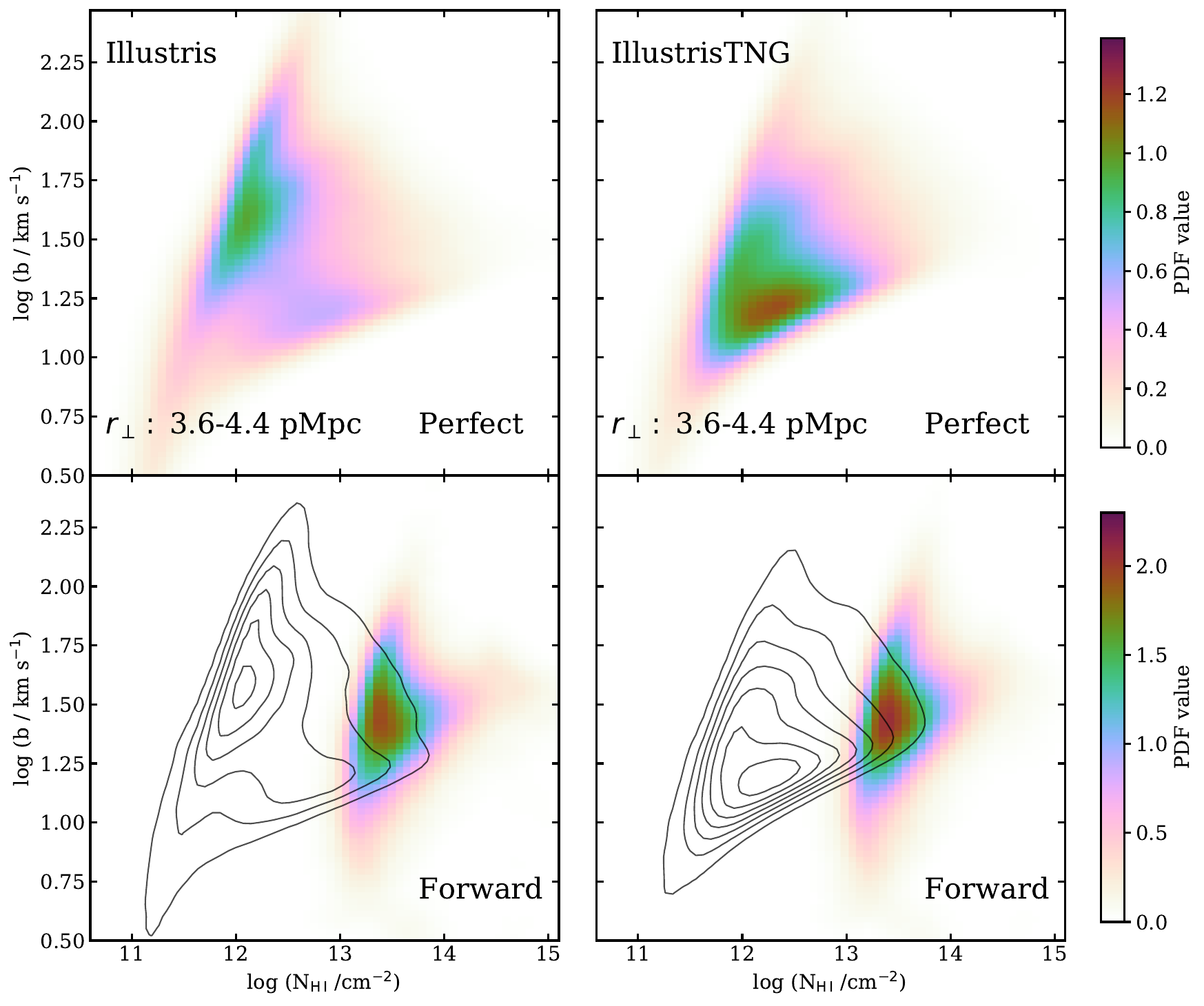}  
\end{subfigure}

\begin{subfigure}{.49\textwidth}
  \centering
  \includegraphics[width=\linewidth,keepaspectratio]{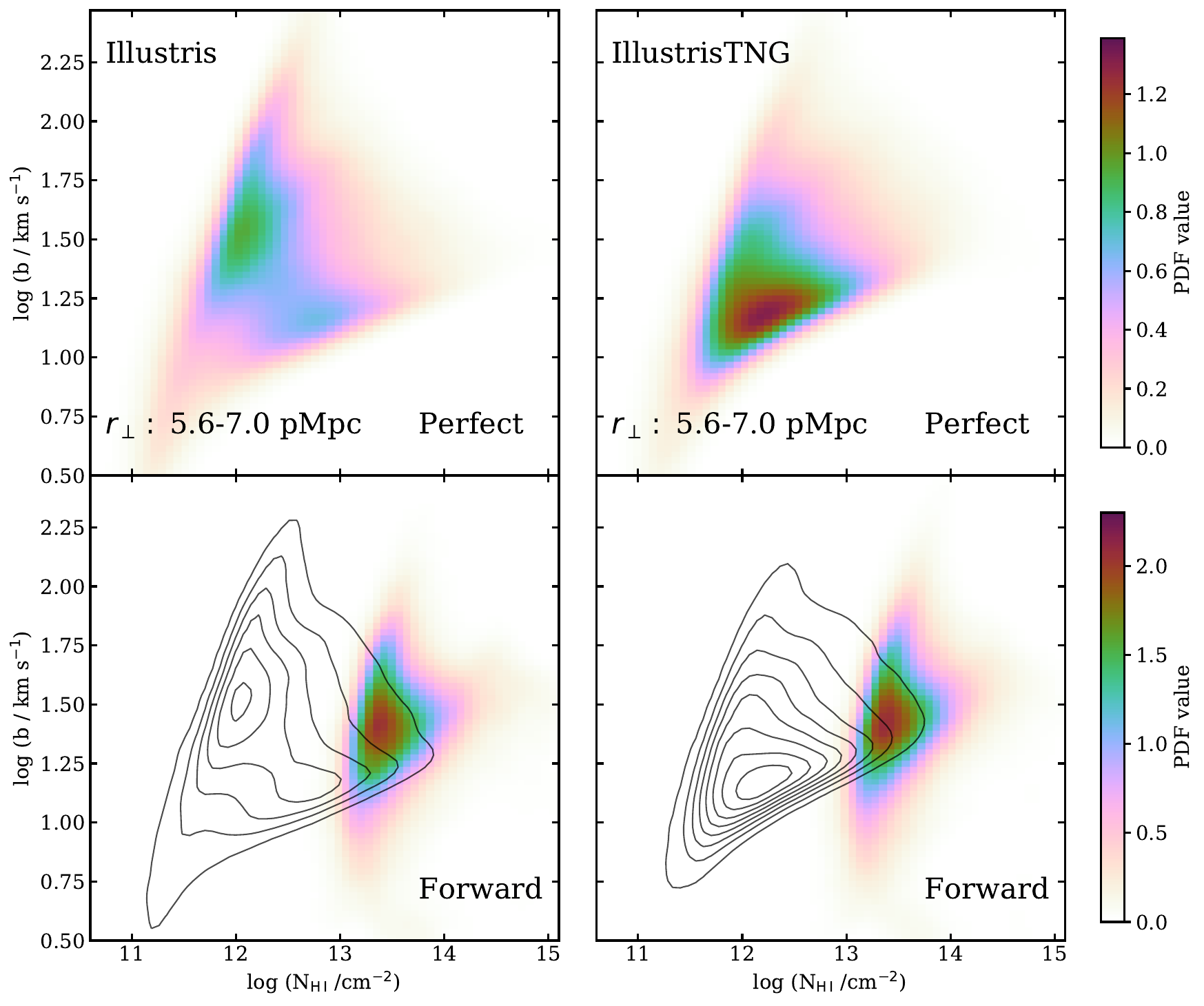}  
\end{subfigure}
\begin{subfigure}{.49\textwidth}
  \centering
  \includegraphics[width=\linewidth,keepaspectratio]{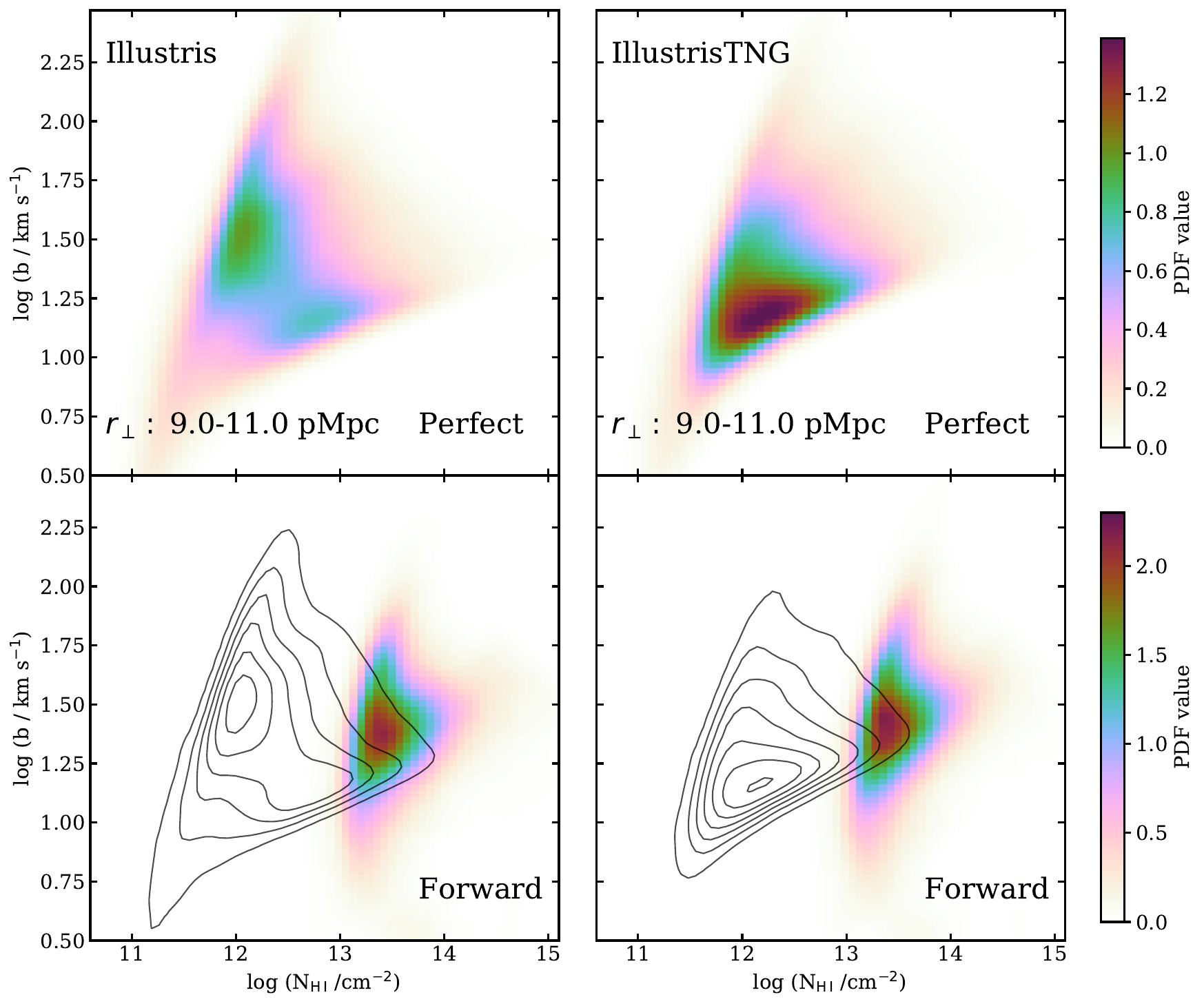}  
\end{subfigure}

\begin{subfigure}{.49\textwidth}
  \centering
  \includegraphics[width=\linewidth,keepaspectratio]{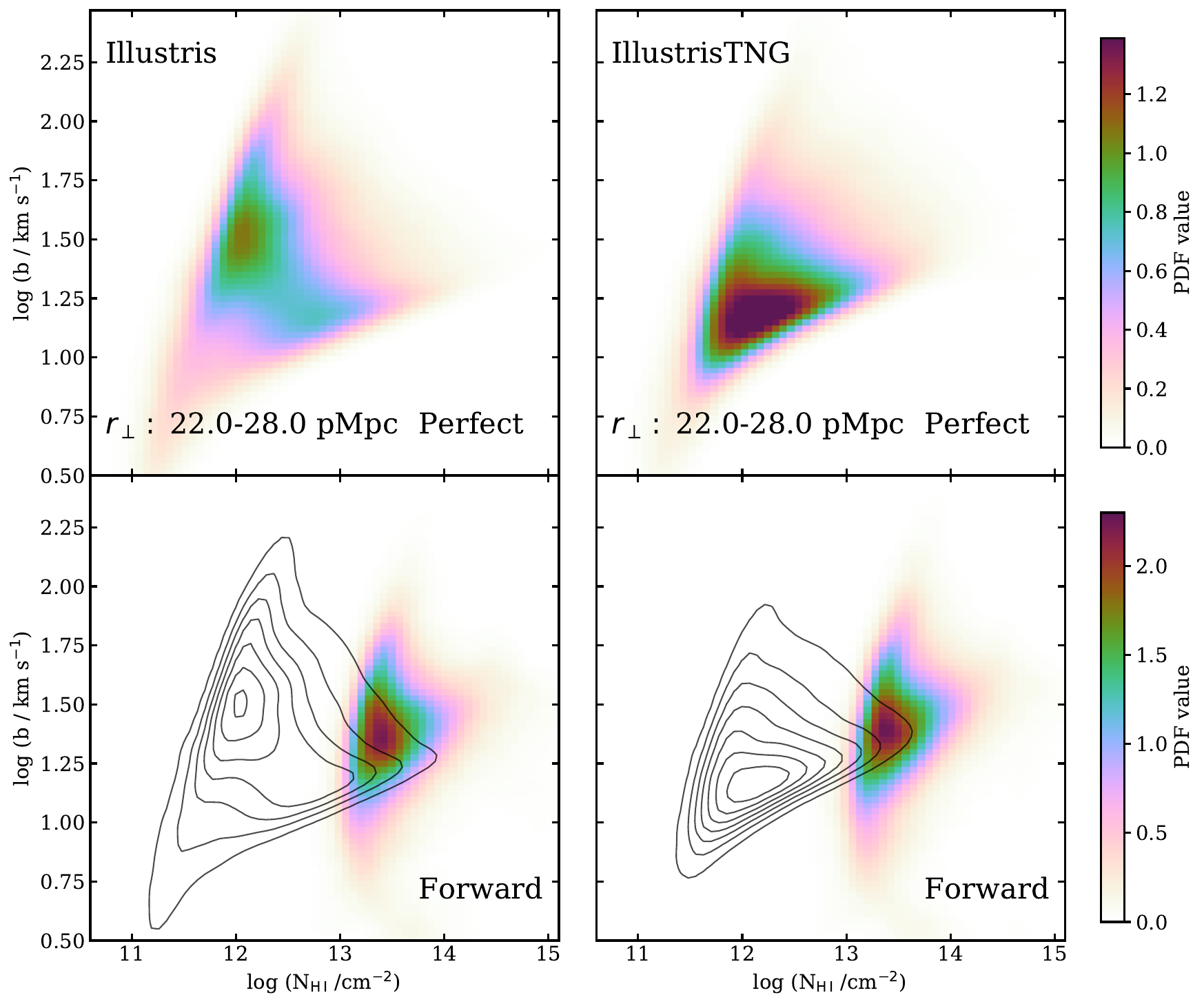}  
\end{subfigure}
\begin{subfigure}{.49\textwidth}
  \centering
  \includegraphics[width=\linewidth,keepaspectratio]{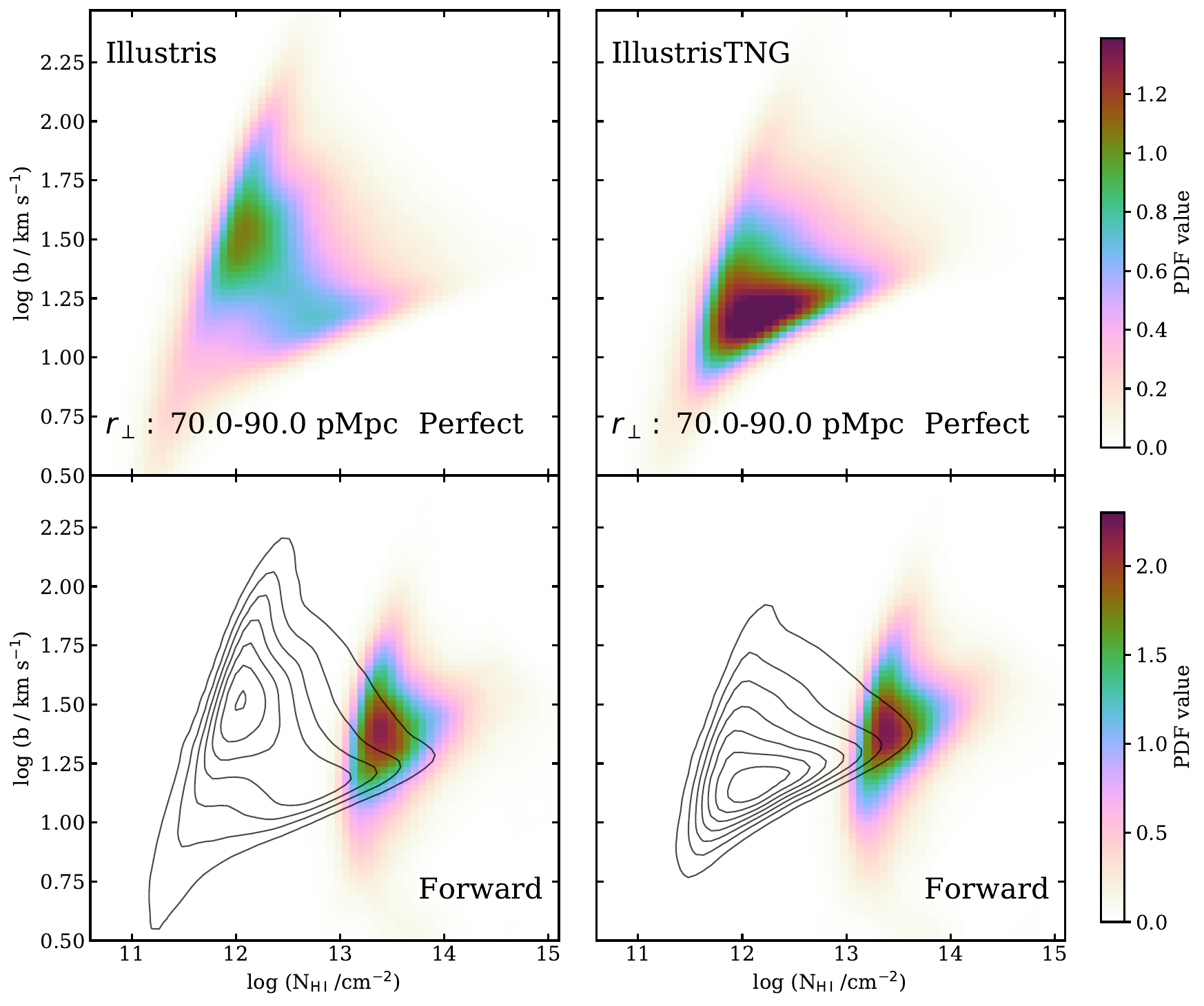}  
\end{subfigure}

\caption{Same as Fig.~\ref{fig.kde_small_impact}, the 2D $b$-\nhi~distribution around massive halos at six different impact parameters (indicated in legends) from 2.2 to 90 pMpc. 
With increasing impact parameters, the shape of the distribution evolves and converges to the one obtained for the IGM (see \citetalias{Khaire23}).}
\label{fig.bn_kde_apendix1}
\end{figure*}

\begin{figure*}
\includegraphics[width=0.98\textwidth,keepaspectratio]{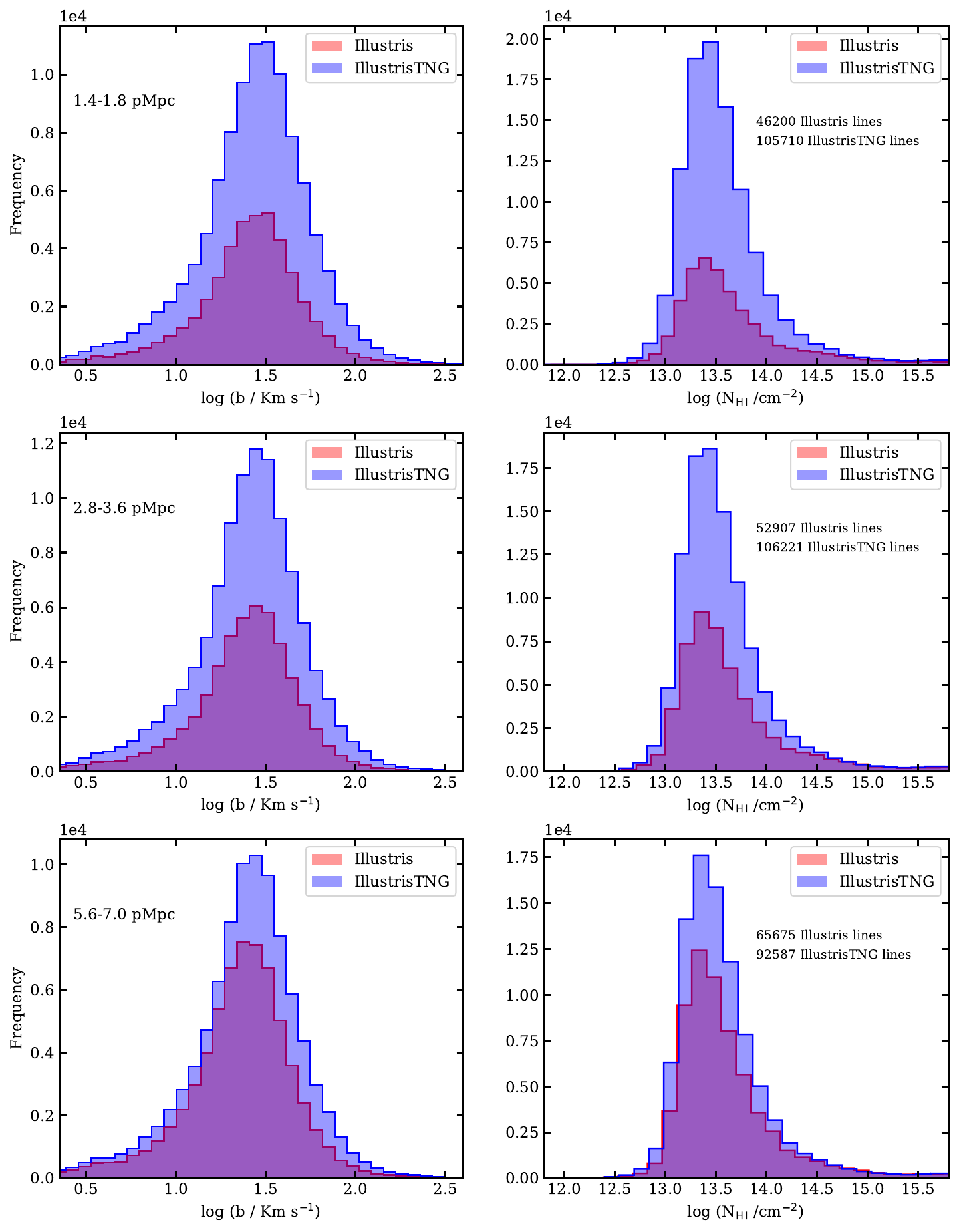}
\caption{The $b$ (right) and \nhi~(left) distribution 
from the forward-modeled sightlines generated at different impact parameter bins
(see legends - left panel) for Illustris and 
IllustrisTNG. The number of lines in each histogram is given in the legend (right-panels). }
\label{fig.halo_forward_hist_1}
\end{figure*}

\begin{figure*}
\includegraphics[width=0.98\textwidth,keepaspectratio]{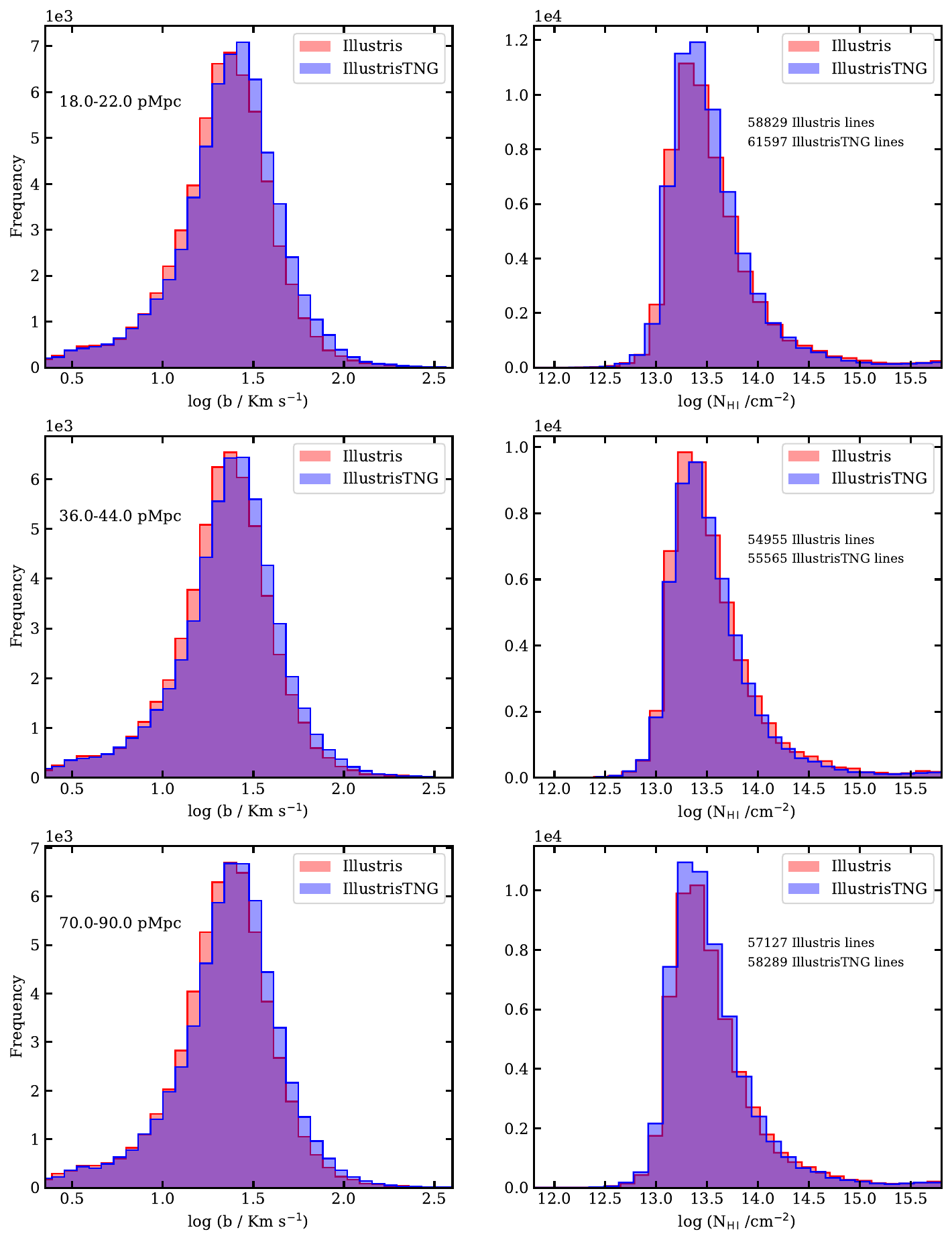}
\caption{Same as Fig.~\ref{fig.halo_forward_hist_1} but for different impact parameters. The number of lines match closely (within $\sim 2$ \%)
for both simulations at high impacts converging to the values for the IGM.}
\label{fig.halo_forward_hist_2}
\end{figure*}

In this section, we show additional figures of 2D and marginalized $b$-\nhi~distributions at 
other impact parameter bins than the one shown in Fig.~\ref{fig.kde_small_impact}, \ref{fig.kde_intermediate_impact}, \ref{fig.kde_large_impact} \& \ref{fig.halo_forward_hist_0}.
The KDE estimated 2D $b$-\nhi~distributions are shown in Fig.~\ref{fig.bn_kde_apendix0} and 
\ref{fig.bn_kde_apendix1} and histograms of $b$ (left-hand panels) and $N_{\rm HI}$ 
(right-hand panels) for forward modeled
sightlines are shown in Fig.~\ref{fig.halo_forward_hist_1} and \ref{fig.halo_forward_hist_2}. See Section~\ref{sec.bn_dist} for the discussion.

\bsp	
\label{lastpage}
\end{document}